\documentclass[manuscript]{emulateapj}

\usepackage{epstopdf}
\usepackage{subfigure}
\usepackage{amsmath}

\usepackage{longtable}

\shorttitle{X-ray Source Heights}
\shortauthors{Reep, Bradshaw, \& Holman}

\begin{document}

\title{X-ray Source Heights in a Solar Flare: Thick-target versus Thermal Conduction Front Heating}

\author{J. W. Reep}
\affil{National Research Council Post-Doc Program, Naval Research Laboratory, Washington, DC 20375, USA}
\email{jeffrey.reep.ctr@nrl.navy.mil}
\author{S. J. Bradshaw}
\affil{Department of Physics and Astronomy, Rice University, Houston, TX 77005, USA}
\email{stephen.bradshaw@rice.edu}
\author{G.D. Holman}
\affil{Solar Physics Laboratory, Code 671, NASA Goddard Space Flight Center, Greenbelt, MD 20771, USA }
\email{gordon.d.holman@nasa.gov}

\begin{abstract}

Observations of solar flares with RHESSI have shown X-ray sources traveling along flaring loops, from the corona down to the chromosphere and back up.  The 28 November 2002 C1.1 flare, first observed with RHESSI by \citet{sui2006} and quantitatively analyzed by \citet{oflannagain2013}, very clearly shows this behavior.  By employing numerical experiments, we use these observations of X-ray source height motions as a constraint to distinguish between heating due to a non-thermal electron beam and {\it in situ} energy deposition in the corona.  We find that both heating scenarios can reproduce the observed light curves, but our results favor non-thermal heating. {\it In situ} heating is inconsistent with the observed X-ray source morphology and always gives a height dispersion with photon energy opposite to what is observed.
\end{abstract}

\keywords{Sun: flares, Sun: X-rays, Sun: corona}

\section{Introduction}

Solar flares are driven by a sudden, explosive release of non-potential magnetic energy, commonly attributed to magnetic reconnection.  Once the energy is released, however, it is not clear how it is partitioned between thermal energy ({\it in situ} heating) and kinetic energy (particle acceleration).  There are two competing theories that individually can explain many (but not all) of the observed features of solar flares.  In one theory, the collisional thick-target model, accelerated electrons are primarily responsible for driving chromospheric evaporation.  In the other, thermal conduction, following {\it in situ} heating of the corona, is the mechanism responsible for filling the post-flare loop.  In this paper, the predictions made by each of these models are tested against observations of X-ray source height evolution in a small C-class flare observed with RHESSI.   

X-ray source heights have previously been used to test models of accelerated electrons in the solar atmosphere.  \citet{brown1975a} showed that X-ray source height measurements can potentially discriminate between the thick- and thin-target models.  Similarly, \citet{brown2002} and \citet{aschwanden2002} used source heights to derive a chromospheric density measurement directly from observed spectra with RHESSI \citep{lin2002}, with a similar analysis performed by \citet{kontar2008} to measure field and density variations with height.

\citet{sui2006} reported an observational measurement of X-ray source heights in a C-class flare observed with RHESSI.  They studied the 28 November, 2002 C1.1 flare (04:35 UT), which occurred near the south-east limb of the Sun.  In the 3-6 keV energy channel, the source was first observed high in the corona before traveling downwards towards the chromosphere, and then rising back into the corona.  Sources observed in two higher energy bands were progressively lower in height than the source in the 3-6 keV band.  Based on RHESSI images and light-curves in multiple energy channels, the authors concluded that the sources were dominated by non-thermal emission, at least during the period of downward motion.  The motion was explained as electrons precipitating downwards with an energy spectrum that hardened with time (causing the source to travel downwards in the atmosphere), which then drove chromospheric evaporation that brought the source back towards the corona by shortening the mean-free path.  

\citet{oflannagain2013} quantitatively analyzed the evolution of this flare, measuring source heights in three different X-ray energy bands with RHESSI (3-6 keV, 6-8 keV, and 8-10 keV).  The sources in all three energy bands were observed to fall from coronal heights towards the chromosphere in the early impulsive phase, before rising back up towards the apex of the loop at later times.  

The authors tested the observed downward motion and height dispersion with energy against the collisional thick-target model \citep{brown1971,emslie1978}, wherein a beam of electrons is accelerated near the loop-top, which then streams through the corona and deposits the bulk of its energy in the upper chromosphere or higher if the density of the loop is sufficiently high, while also emitting hard X-rays (HXRs) via non-thermal bremsstrahlung.  As the electrons deposit their energy through Coulomb collisions with the ambient ions, the pressure in the chromosphere rises, which drives an expansion of material back into the corona (termed chromospheric evaporation by \citealt{hirayama1974}, also described in the important works of \citealt{sweet1969, brown1973, antiochos1978, fisher1985a,fisher1985b}).  This fills the coronal part of the loop with hot plasma that causes it to light up in the soft X-ray (SXR) bands.  

There is another possibility: the loop is heated {\it in situ} near its apex, driving a thermal conduction front through the corona towards the foot-points, heating material to very high temperatures as it propagates.  In this way, the X-ray sources might also be expected to follow a trend towards the chromosphere, and then rise back up into the corona as heated material is evaporated from the lower atmosphere.    

In this paper, the observational measurements of RHESSI source heights are tested against both heating mechanisms using a state-of-the-art hydrodynamic model.  Numerical experiments of the flare are performed using both the thick-target model and the thermal conduction model.  X-ray spectra are calculated as a function of time and position in the model, and height measurements are synthesized for the RHESSI energy bands with the same techniques used to derive the observed heights.  The synthesized heights are then compared to the results of \citet{oflannagain2013}, to contrast the predictions of the different heating mechanisms and to determine which, if any, are consistent with the observed motions of the sources.  

In Section \ref{modeling}, the methodology and numerical modeling is briefly discussed, and the method by which heights are calculated is explained.  In Section \ref{results}, the results of a large number of simulations of the flare using the different heating mechanisms are presented.  Finally, in Section \ref{conclusions}, the results are summarized and interpreted.  

\section{Numerical Modeling}
\label{modeling}

The numerical experiments were performed with the HYDRAD code \citep{bradshaw2013}, which solves the hydrodynamic equations that describe the conservation of mass, momentum, and energy for an isolated magnetic flux tube and a multi-fluid plasma ({\it e.g.} electrons, ions, neutrals).  The equations and assumptions are explained in the appendix of \citet{bradshaw2013}.  As detailed in \citet{reep2013}, HYDRAD has been modified to include thick-target heating due to a non-thermal population of electrons, taking account of the non-uniform ionization structure along the path of propagation, as well as using a more realistic chromosphere, based on the VAL C model \citep{vernazza1981}, with a non-uniform ionization structure.  The effects of neutral atoms on the energy balance are treated in full (hydrogen ionization/recombination and thermal conduction), while the radiation treatment is based on the recipes derived by \citet{carlsson2012}.  The loops are assumed to be semi-circular, along the field-aligned direction, with a constant cross-sectional area.  The initial transition region and coronal temperature and density profiles are found by integrating the hydrostatic equations from the top of the VAL chromosphere to the apex of the coronal loop.  The electron and ion populations are assumed to be in thermal equilibrium before any heating occurs.  

All of the experiments were performed for a loop of length $2L = 30$\,Mm (which was estimated from the observed loop).  The cross-sectional area estimated from the observed RHESSI sources was approximately $2.6 \times 10^{17}$\,cm$^{2}$, which we take as an upper limit.  The cross-sectional area is assumed constant across the loop, since the observational source widths did not indicate a non-uniform structure.  However, \citet{emslie1992} modeled loops with apex to foot-point cross-sectional area ratios $R$ ranging from 1 to 30, and found two important results.  First, the electrons did not penetrate as deeply into the chromosphere for large $R$, so that heating was more strongly confined to the top of the chromosphere.  Second, the induced evaporation velocities were much higher for large $R$ because the heating is higher in the chromosphere (compare \citealt{reep2015}).  If the cross-sectional area were not assumed constant, the up-flows, and thus upward source motions, would travel more quickly. 

The initial density and temperature profiles are calculated to be consistent with the density found by \citet{oflannagain2013}.  The initial loop profile has a coronal temperature of around 4\,MK and density around $4 \times 10^{10}$\,cm$^{-3}$.  Lower initial density values were found to be inconsistent with the observations, regardless of the heating mechanism, due to the lengthened mean-free paths of accelerated electrons and reduction in thermal emission.  The loop was observed to have a thermal source near its apex at the onset of the flare ($\gtrsim 10$\,MK), so it is preheated with a short thermal burst lasting 1 second in all of the experiments.

To compare and contrast the heating mechanisms, numerical experiments have been performed with two forms of energy input.  In the thick-target model, we assume an electron distribution of the form
\begin{equation}
\mathfrak{F}(E_{0}, t) = \frac{F_{0}(t)}{E_{c}^{2}}\ (\delta - 2) \times
  \begin{cases}
   0 & \text{if } E_{0} < E_{c} \\
   \Big(\frac{E_{0}}{E_{c}}\Big)^{- \delta}       & \text{if } E_{0} \geq E_{c}
  \end{cases}
\label{sharpdist}
\end{equation}

\noindent which is generally referred to as a sharp cut-off.  $F_{0}(t)$ is the energy flux carried by the electron beam (erg s$^{-1}$ cm$^{-2}$), $E_{c}$ is the low-energy cut-off (erg), $\delta$ is the spectral index of the power-law, and $E_{0}$ is the initial energy of a given electron in the beam (erg).

We use the heating functions derived by \citet{emslie1978}, for a beam of electrons depositing their energy via Coulomb collisions with a hydrogen target.  Following the approach of \citet{hawley1994}, we generalize Emslie's solution for a non-uniform ionization structure, which is important for recovering spectral breaks in observed X-ray spectra \citep{kontar2003}.  See \citet{holman2011} for an overview of electron distributions in the thick-target model.  In the experiments with a non-thermal component, the time-dependent spectral index $\delta$ found in Figure 1a of \citet{oflannagain2013} is used (initially around 7.5, quickly hardening to 4.1, and then gradually softening again).  

In the thermal conduction model, the heat is deposited {\it in situ} in the corona, causing a thermal conduction front to propagate towards the chromosphere.  We assume the temporal envelope of the heating profile is triangular (but not necessarily symmetric about the time of peak heating), with a total duration of 100 seconds of heating.  The energy is deposited over a length scale of [1, 10, 30]\,Mm, centered at the apex of the loop.  

X-ray emissions are calculated in full, over the entire loop and for each time step.  Thermal emissions, including line emissions, thermal free-free, thermal free-bound, and two-photon continua, are calculated using CHIANTI v.8 over the wavelength range of interest \citep{dere1997,delzanna2015}.  Non-thermal free-free emission is calculated directly by evaluating the standard bremsstrahlung integral (see the appendix of \citealt{reep2013}), using a fully relativistic cross-section \citep{bethe1934}. 
\LTcapwidth=\textwidth

\begin{deluxetable*}{c c c c c c c c c c c}
\caption{The results of various numerical experiments of the flare.  The following are listed: the run number, along with the heating type, temporal envelope (rise and fall times in seconds), non-thermal energy flux $F_{NT}$ (erg s$^{-1}$ cm$^{-2}$), thermal heat input $H_{Th}$ (erg s$^{-1}$ cm$^{-3}$), cross-sectional area (cm$^{2}$), total energy input $E_{total}$ (erg), the length over which {\it in situ} heating was deposited (Mm), the cut-off energy (keV), and finally the synthesized emission in the two GOES channels (W m$^{-2}$).  Note that the observed (background subtracted) GOES flux was B3.8/A4.8. \label{20021128sim}} \\
\hline 
Run & Type & T & $F_{NT}$ & $H_{Th}$ &  $A$ & $E_{Total}$ & Heat L & $E_{c}$ & 1-8 \AA\ & 0.5-4 \AA\ \\
\hline
1 & Non-Thermal & 50-50 & $1.66 \times 10^{10}$ & - & $1.0 \times 10^{17}$ & $8.3 \times 10^{28}$ & - & 1 & B3.8 & A2.5 \\
2 & Non-Thermal & 50-50 & $3.20 \times 10^{9}$ & - & $1.0 \times 10^{17}$ & $1.6 \times 10^{28}$ & - & 3 & B3.8 & A3.5 \\
3 & Non-Thermal & 50-50 & $2.40 \times 10^{9}$ & - & $1.0 \times 10^{17}$ & $1.2 \times 10^{28}$ & - & 5 & B3.8 & A5.6 \\
4 & Non-Thermal & 50-50 & $2.00 \times 10^{9}$ & - & $1.0 \times 10^{17}$ & $1.0 \times 10^{28}$ & - & 7 & B3.8 & A6.6 \\
5 & Non-Thermal & 50-50 & $6.54 \times 10^{10}$ & - & $2.6 \times 10^{16}$ & $8.5 \times 10^{28}$ & - & 1 & B3.8 & A3.2 \\
6 & Non-Thermal & 50-50 & $1.15 \times 10^{10}$ & - & $2.6 \times 10^{16}$ & $1.5 \times 10^{28}$ & - & 3 & B3.8 & A3.7 \\
7 & Non-Thermal & 50-50 & $1.00 \times 10^{10}$ & - & $2.6 \times 10^{16}$ & $1.3 \times 10^{28}$ & - & 5 & B3.8 & A6.1 \\
8 & Non-Thermal & 50-50 & $9.23 \times 10^{9}$ & - & $2.6 \times 10^{16}$ & $1.2 \times 10^{28}$ & - & 7 & B3.8 & A7.8 \\
9 & Non-Thermal & 50-50 & $1.78 \times 10^{11}$ & - & $1.0 \times 10^{16}$ & $8.9 \times 10^{28}$ & - & 1 & B3.8 & A4.0 \\
10 & Non-Thermal & 50-50 & $2.20 \times 10^{10}$ & - & $1.0 \times 10^{16}$ & $1.1 \times 10^{28}$ & - & 3 & B3.8 & A3.1 \\
11 & Non-Thermal & 50-50 & $1.90 \times 10^{10}$ & - & $1.0 \times 10^{16}$ & $9.5 \times 10^{27}$ & - & 5 & B3.7 & A5.0 \\
12 & Non-Thermal & 50-50 & $1.86 \times 10^{10}$ & - & $1.0 \times 10^{16}$ & $9.3 \times 10^{27}$ & - & 7 & B3.8 & A6.3 \\
\hline
13 & Thermal & 50-50 & - & 25 & $2.7 \times 10^{16}$ & $3.4 \times 10^{27}$ & 1 & - & B3.8 & A2.1 \\
14 & Thermal & 50-50 & - & 50 & $1.8 \times 10^{16}$ & $4.5 \times 10^{27}$ & 1 & - & B3.8 & A2.3 \\
15 & Thermal & 50-50 & - & 100 & $1.0 \times 10^{16}$ & $5.0 \times 10^{27}$ & 1 & - & B3.8 & A2.6 \\
16 & Thermal & 50-50 & - & 250 & $3.4 \times 10^{15}$ & $4.3 \times 10^{27}$ & 1 & - & B3.8 & A3.0 \\
17 & Thermal & 50-50 & - & 500 & $1.2 \times 10^{15}$ & $3.1 \times 10^{27}$ & 1 & - & B3.8 & A3.5 \\
18 & Thermal & 25-75 & - & 25 & $2.8 \times 10^{16}$ & $3.5 \times 10^{27}$ & 1 & - & B3.8 & A2.1 \\
19 & Thermal & 25-75 & - & 50 & $1.9 \times 10^{16}$ & $4.7 \times 10^{27}$ & 1 & - & B3.8 & A2.2 \\
20 & Thermal & 25-75 & - & 100 & $1.1 \times 10^{16}$ & $5.5 \times 10^{27}$ & 1 & - & B3.8 & A2.5 \\
21 & Thermal & 25-75 & - & 250 & $3.5 \times 10^{15}$ & $4.4 \times 10^{27}$ & 1 & - & B3.8 & A2.8 \\
22 & Thermal & 50-50 & - & 25 & $4.5 \times 10^{15}$ & $5.6 \times 10^{27}$ & 10 & - & B3.8 & A2.7 \\
23 & Thermal & 50-50 & - & 50 & $1.7 \times 10^{15}$ & $4.2 \times 10^{27}$ & 10 & - & B3.8 & A3.3 \\
24 & Thermal & 50-50 & - & 25 & $3.3 \times 10^{15}$ & $1.2 \times 10^{28}$ & 30 & - & B3.8 & A2.9 \\
\hline
\end{deluxetable*}

Non-thermal free-bound emission has also been included for completeness \citep{brown2008,brown2009,brown2010}, and may contribute to the emission within the regime considered in this work, particular for electron beams having lower energy cut-offs.  Recent work \citep{warren2014} suggests that flares have photospheric abundances, so we choose to use the abundance set of \citet{asplund2005}.  \citet{brown2009} note that the cross-sections for recombination scale as $A_{Z} Z^{4}$ where $A_{Z}$ is the abundance of an ion with charge $Z$.  With this abundance set, iron is the most important element ($A_{Z} Z^{4} \approx 14.5$ for Fe XXVII), while oxygen, the next most important element, is essentially negligible ($A_{Z} Z^{4} \approx 2.0$ for O IX).  We thus only include contributions from iron ions hotter than 10 MK (Fe XXI and above).  

RHESSI source heights are calculated as in \citet{oflannagain2013}, as follows.  Note that \citet{oflannagain2013} use the word ``height'' to refer to the field-aligned coordinate.  For the sake of consistency, the word ``height'' in this paper is used the same way.  In common with those authors, all flux below 30\% of the brightest emission is removed to isolate true sources from background emission.  Emission is ignored entirely if the maximum is below $10^{-2}$\,photons s$^{-1}$ cm$^{-2}$ keV$^{-1}$ as RHESSI is not able to detect such small intensities.  The emission is integrated over 8 seconds (except the first height calculation which was integrated over 4 seconds), with each data point determined at a 4 second cadence (which means there is an overlap between adjacent points).  The emission is calculated in three energy channels: 3-6 keV, 6-8 keV, and 8-10 keV.

In the numerical model, the X-ray emissions are calculated at the spatial resolution of the code, which is much finer than RHESSI.  To correct for this, the grid cells are binned onto a coarser, uniform grid of 2 arcsecond cells, which is approximately RHESSI's pixel size.  In this case, the flare occurred at the eastern limb of the disk, so projection effects are small (in comparison, for example, with the methodology of \citealt{bradshaw2011}, which calculates emission assuming the loop is at disk center).  Additionally, albedo is least significant near the solar limb, with the fraction of reflected photons quickly falling to zero \citep{tomblin1972,santangelo1973,kontar2011}, so that, for this flare, it should not have altered the measured spectra, nor should it impact the results significantly.

\citet{oflannagain2013} calculate their source heights under the assumption that the sources are 2-dimensional Gaussian structures.  The fits to the sources determine the centroids of emission, along a curve that passes through both foot-points.  \citet{sui2006} measured the source heights using the centroid of the 60\% contours, which does not assume a particular shape to the emission.  Since the numerical experiments are 1-dimensional, and since the calculated emission often is not Gaussian, we similarly calculate the source height centroid more directly: 
\begin{equation}
H = \frac{\sum\limits_{i=1}^{N}\Big(x_{i}\ \frac{I_{i}}{I_{max}}\Big)}{\sum\limits_{i=1}^{N}\frac{I_{i}}{I_{max}}}
\end{equation}
\noindent where $x_{i}$ is the position and $I_{i}$ is the intensity of a given data point, $I_{max}$ is the maximum of the intensity, and $N$ is the total number of data points.  There is one small caveat that must be accounted for.  At early times in the thick-target simulations before significant thermal emission develops, there are two sources of emission, one at each foot-point, that must be fit (compare Figure 1c of \citealt{oflannagain2013}).  Two centroids are therefore calculated, although we choose to follow the lower leg of the loop, as done by the observers.

\section{Results}
\label{results}

\begin{figure*}[H]
\begin{minipage}[t]{0.5\linewidth}
\centering
\includegraphics[width=3.3in]{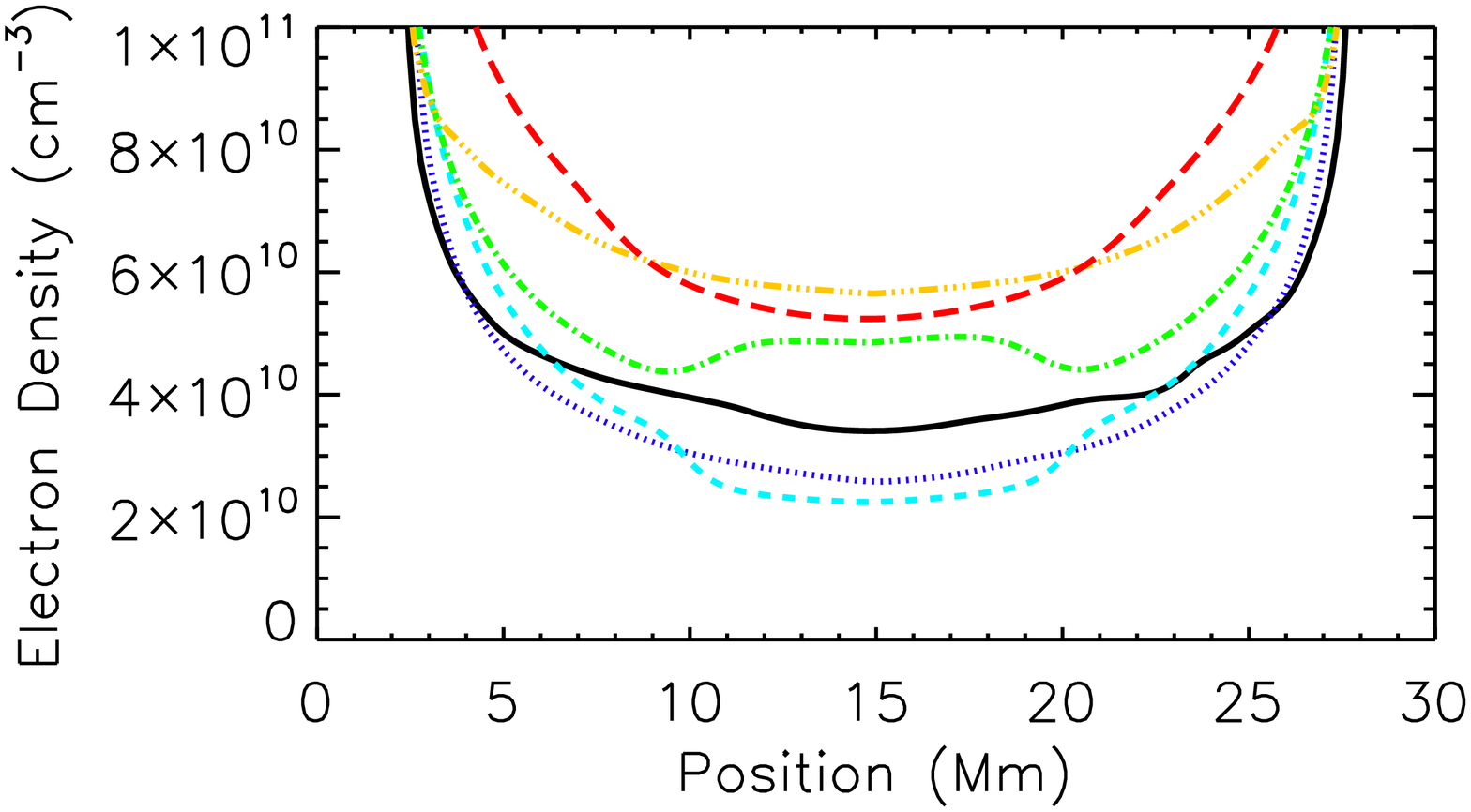}
\end{minipage}
\begin{minipage}[t]{0.5\linewidth}
\centering
\includegraphics[width=3.3in]{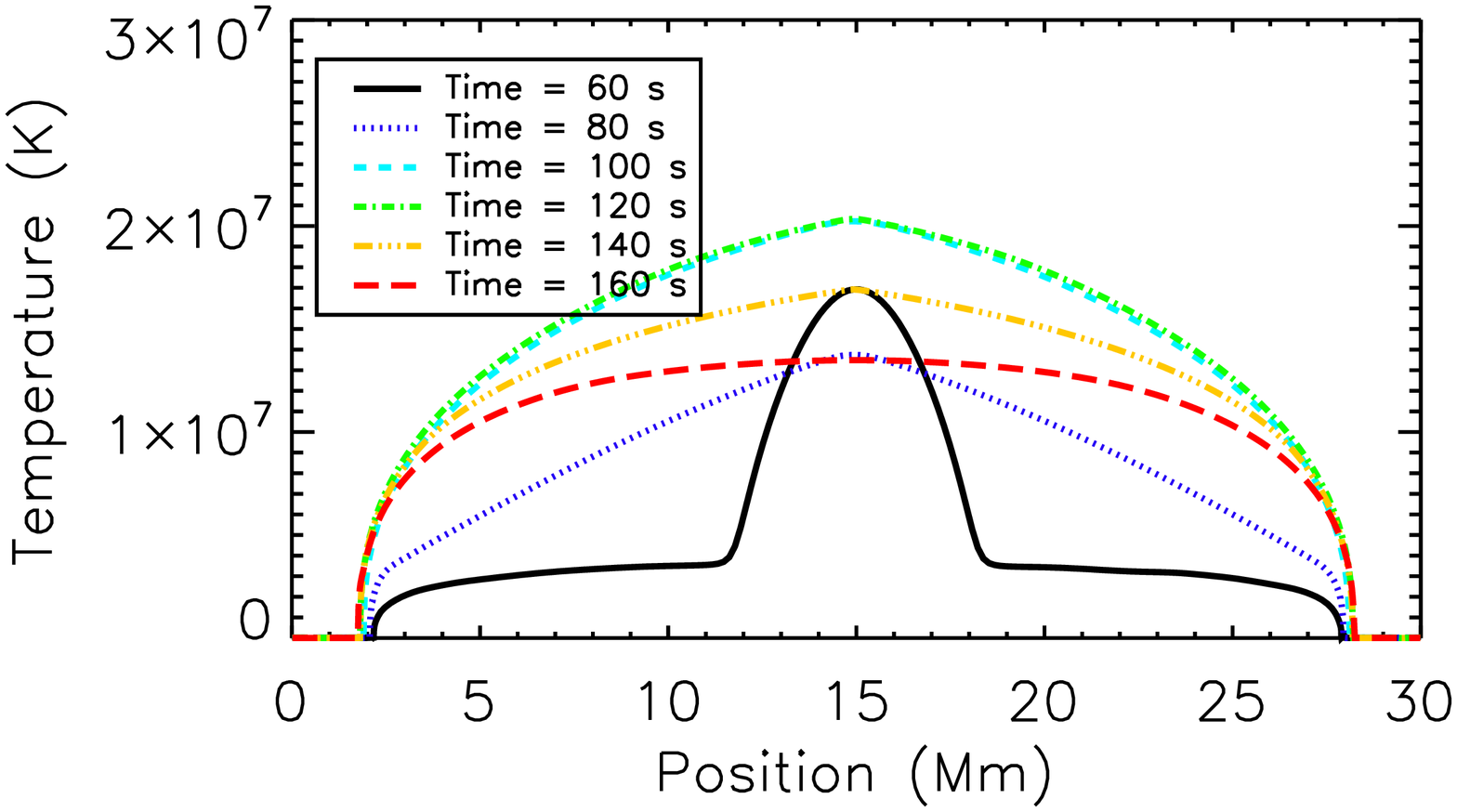}
\end{minipage}
\begin{minipage}[b]{0.5\linewidth}
\centering
\includegraphics[width=3.3in]{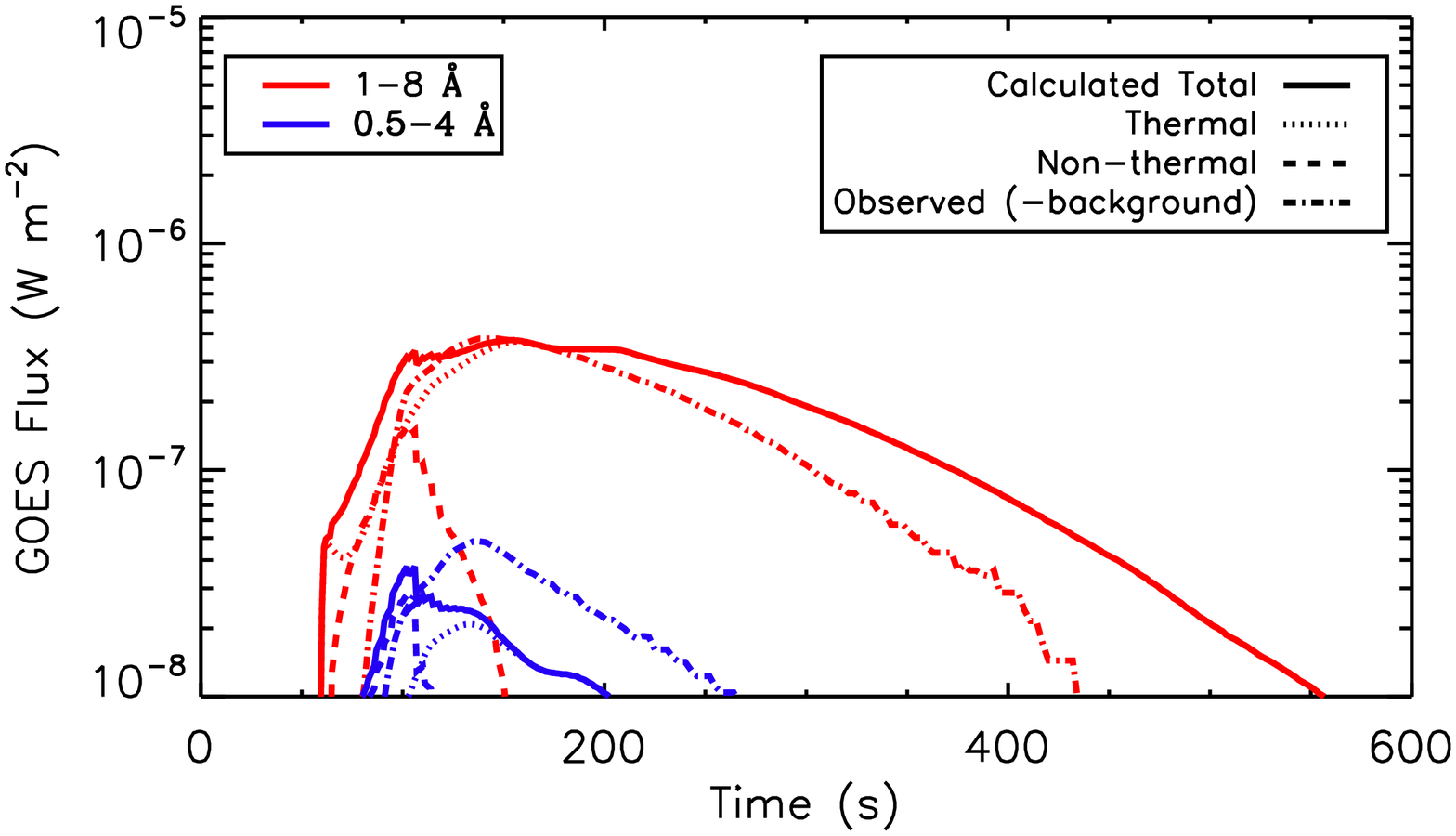}
\end{minipage}
\begin{minipage}[b]{0.5\linewidth}
\centering
\includegraphics[width=3.3in]{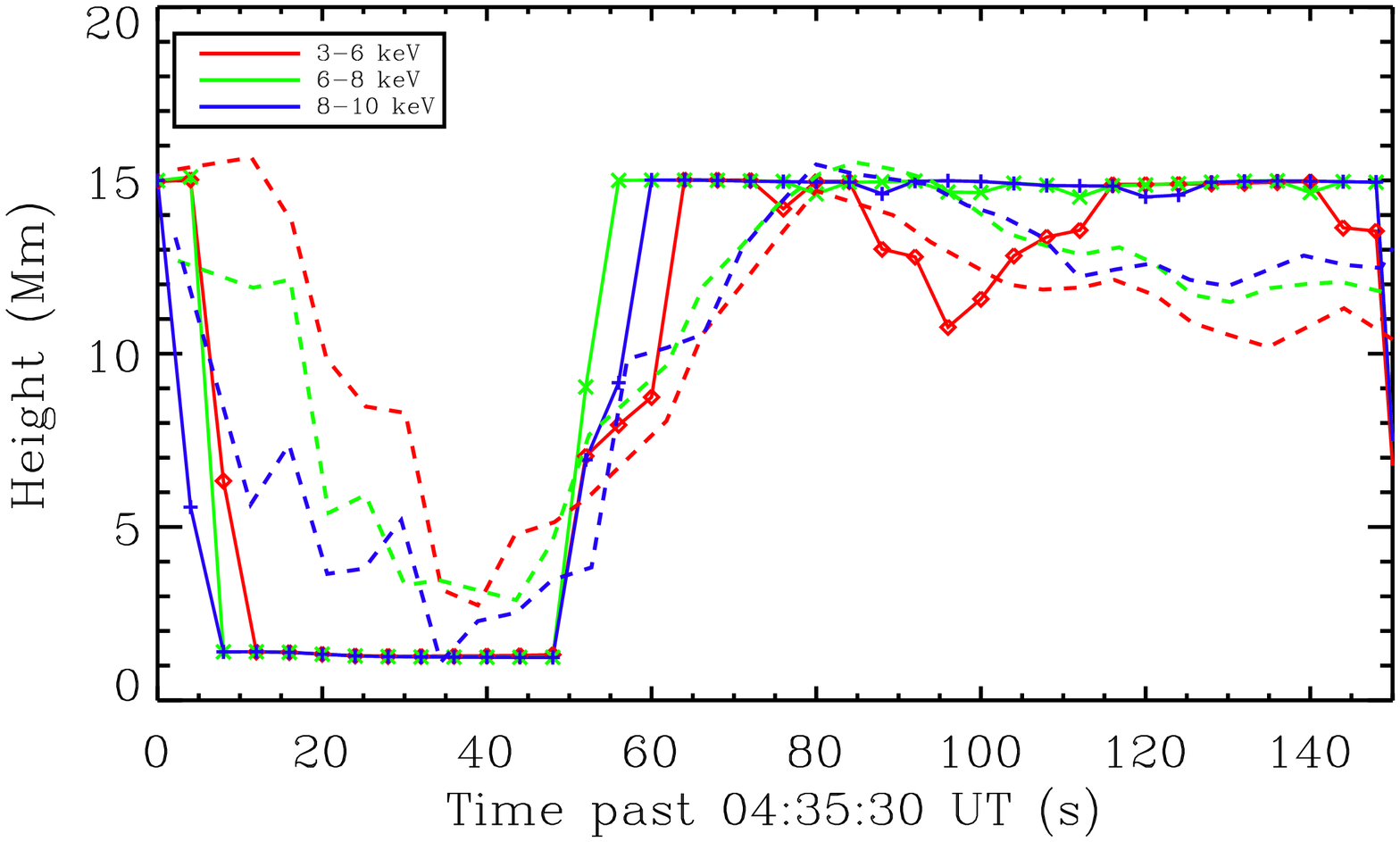}
\end{minipage}
\caption{The results for Run 6, heated by a beam of electrons with a total energy of $1.5 \times 10^{28}$\,erg.   {\it Top Left:} The electron density vs position along the loop, with several times over-plotted.  {\it Top Right:} The electron temperature versus position along the loop. {\it Bottom Left:} The synthesized GOES light-curve, with the observed (background-subtracted) light-curve overlaid.  {\it Bottom Right:} The calculated source heights for the three energy bands (3-6, 6-8, 8-10 keV, solid lines), with the observed heights overlaid (dashed lines; see also Figure 2 of \citealt{oflannagain2013}).   }
\label{run6}
\end{figure*}
\begin{figure*}[H]
\begin{minipage}[b]{0.5\linewidth}
\centering
\includegraphics[width=3.3in]{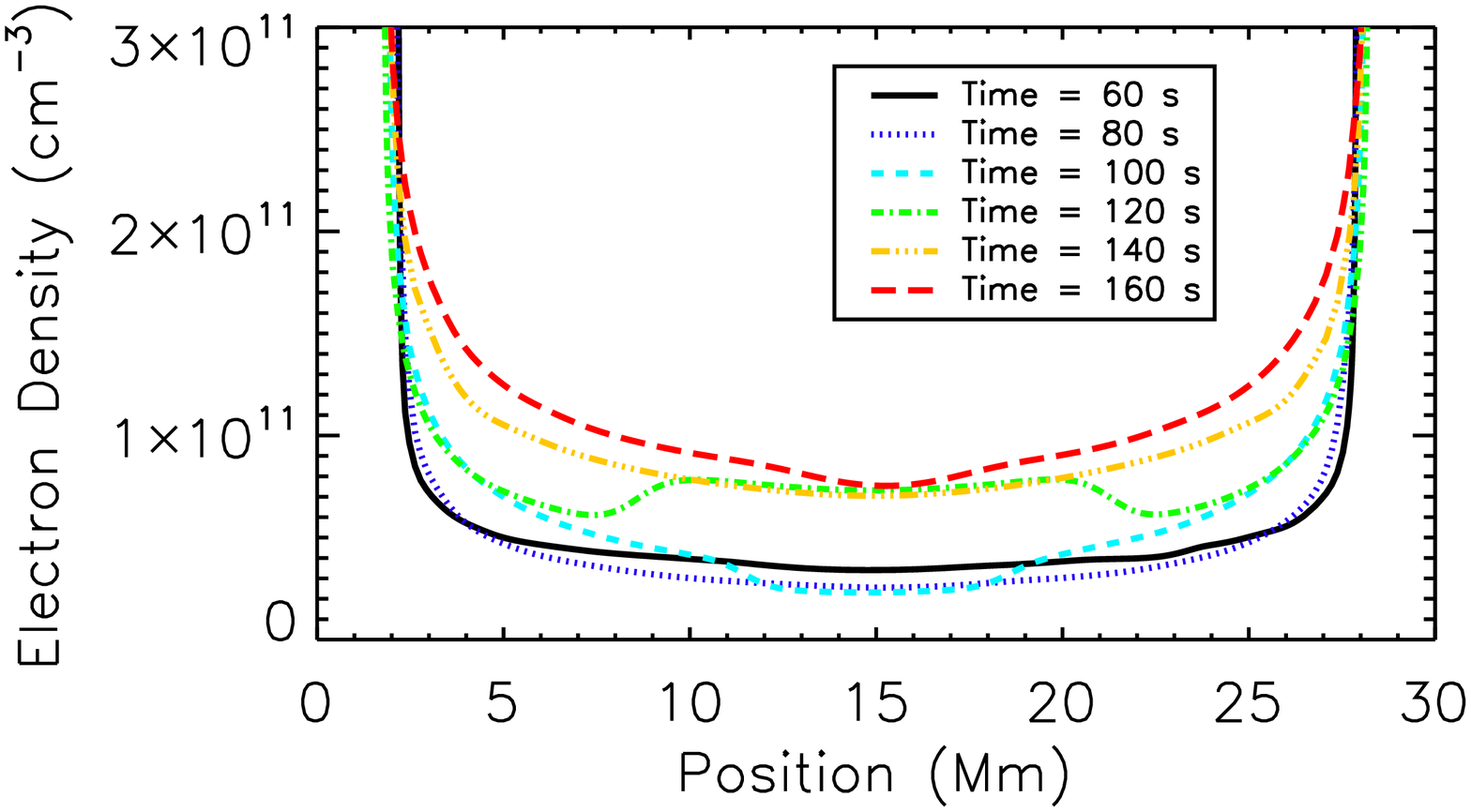}
\end{minipage}
\begin{minipage}[b]{0.5\linewidth}
\centering
\includegraphics[width=3.3in]{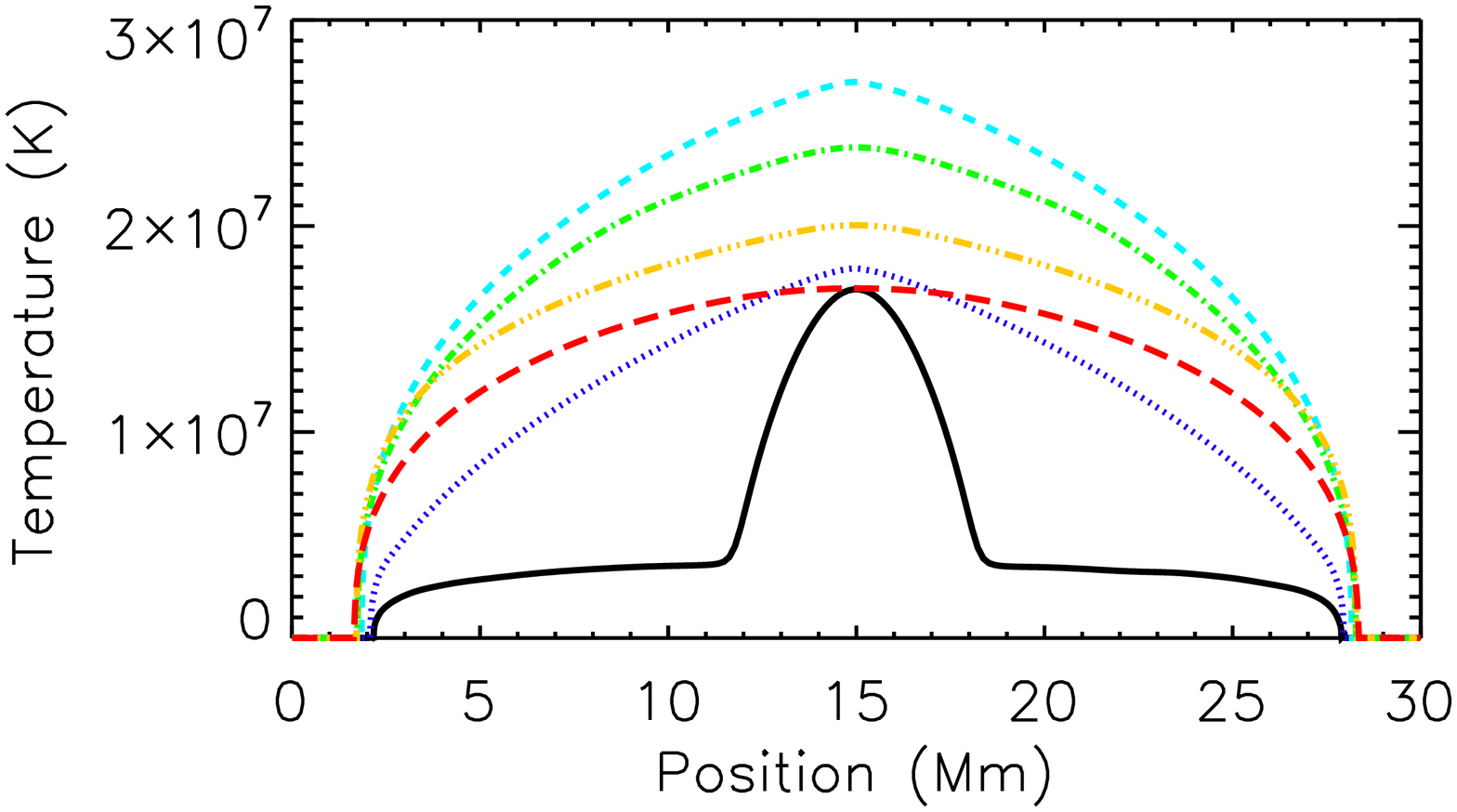}
\end{minipage}
\begin{minipage}[b]{0.5\linewidth}
\centering
\includegraphics[width=3.3in]{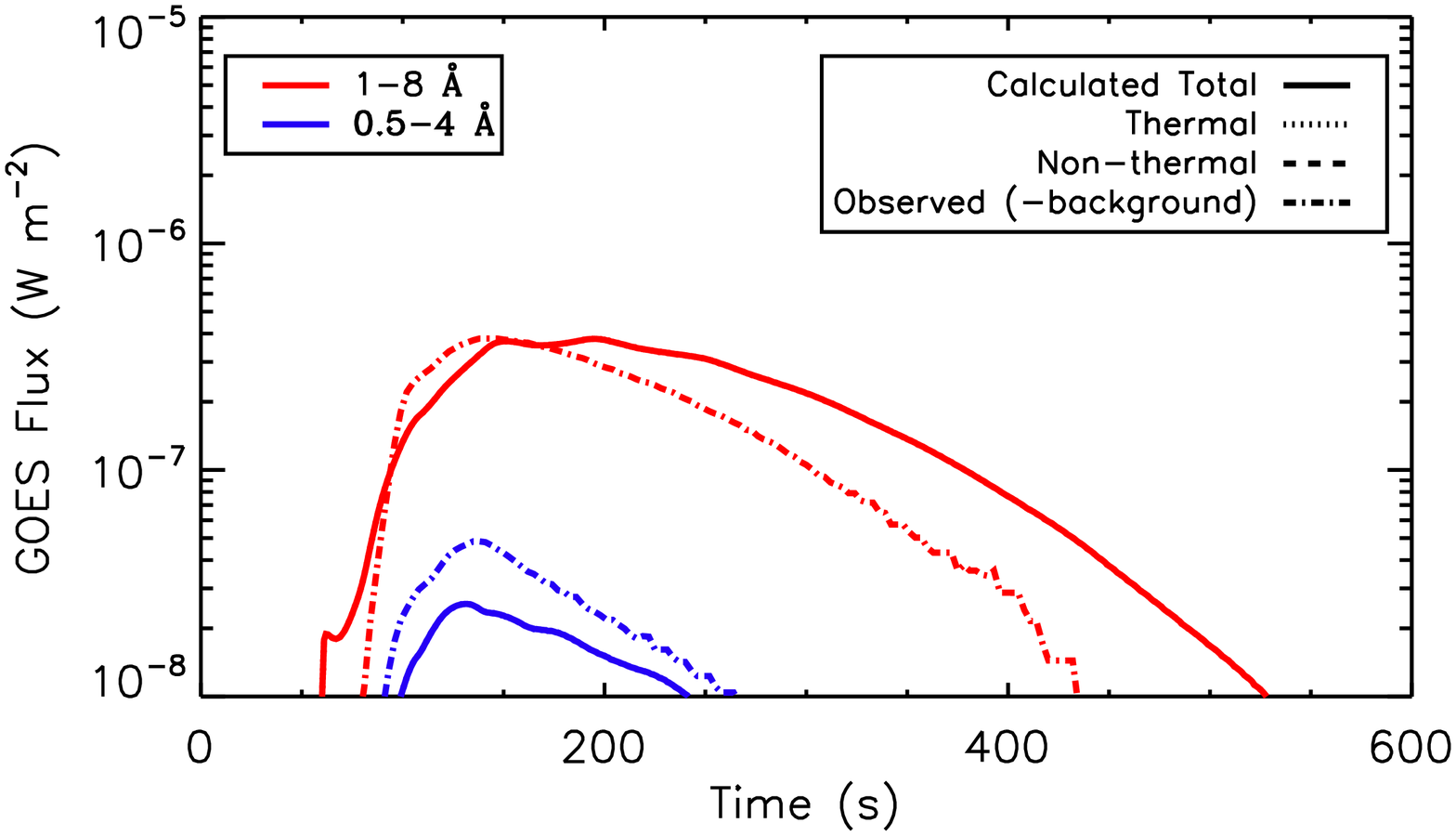}
\end{minipage}
\begin{minipage}[b]{0.5\linewidth}
\centering
\includegraphics[width=3.3in]{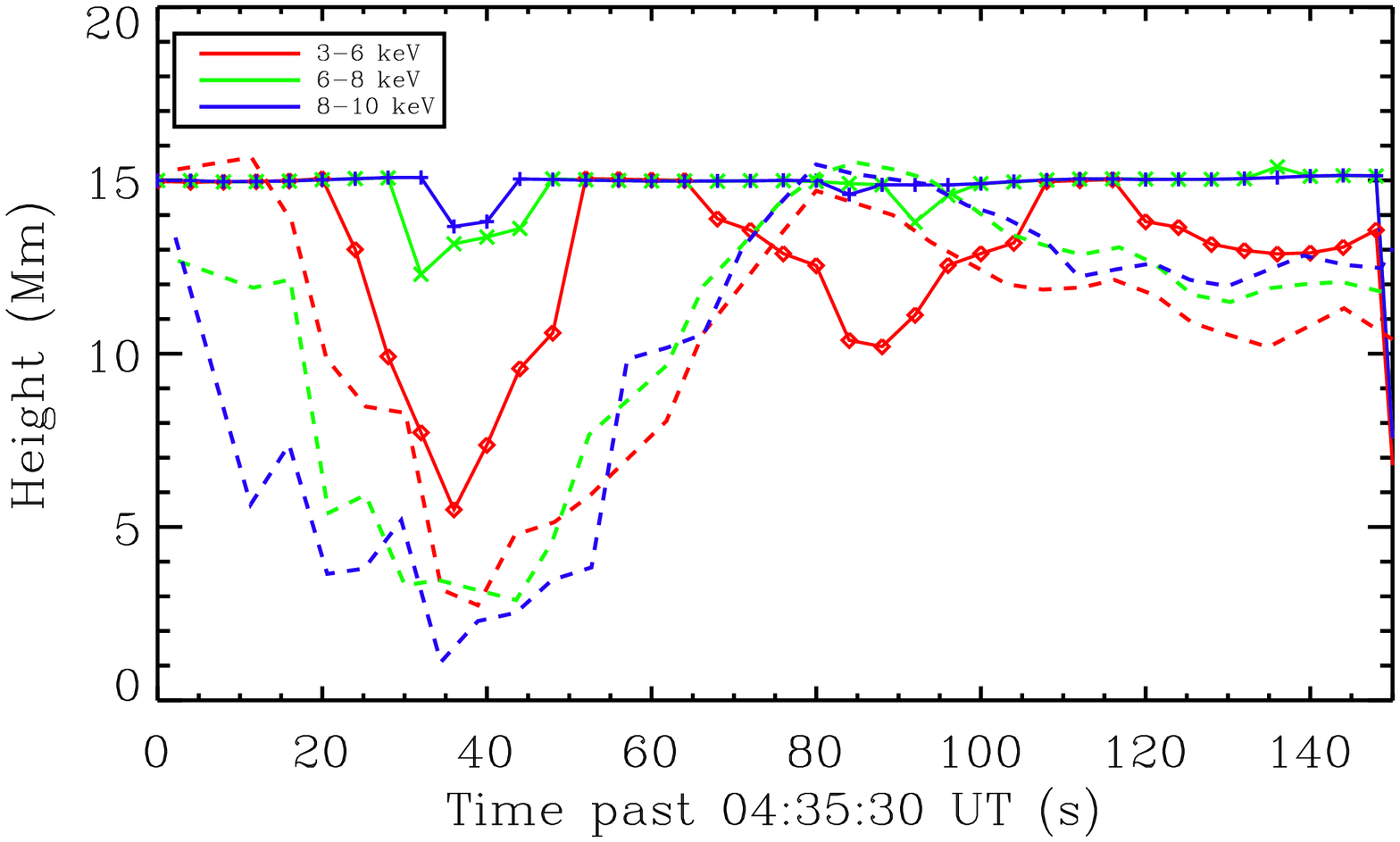}
\end{minipage}
\caption{The results for Run 15, heated {\it in situ} over 1\,Mm near the apex of the loop with a total energy of $5.0 \times 10^{27}$\,erg.  The plots are as in Figure \ref{run6}.   }
\label{run15}
\end{figure*}

24 numerical experiments have been performed, with parameters listed in Table \ref{20021128sim}.  Twelve experiments have been performed with a thick-target model, all of which used a beam that lasted for 100 seconds, with a 50 second rise and 50 second decay time.  The total non-thermal energy was varied in these runs to best approximate the GOES class in the 1-8 \AA\ channel.  Note that the observed flare, after background subtraction, had a peak GOES flux of B3.8 (1-8 \AA) and A4.8 (0.5-4 \AA).  The cut-off energy was varied between [1, 3, 5, 7]\,keV, since the observational value was uncertain (\citealt{oflannagain2013} note that the non-thermal emission extends down to RHESSI's observational limit of 3 keV).  Recall that the spectral index $\delta$ was time-dependent, so that the electron beam changes in time ({\it i.e.} the spectrum hardened from about 7.5 to 4.1 and subsequently softened).  12 experiments adopted {\it in situ} coronal heating, over a length scale of [1, 10, 30]\,Mm centered at the loop apex, for durations of 100 seconds.  The time profiles of heating were triangular, with rise and fall times of 50 seconds each, or 25 and 75 seconds.

Consider Run 6, which was heated with a non-thermal electron beam, carrying total non-thermal energy $1.5 \times 10^{28}$\,erg (corresponding to a maximal beam flux of $1.15 \times 10^{10}$\,erg s$^{-1}$ cm$^{-2}$).  The cut-off energy was assumed to be 3\,keV.  Figure \ref{run6} shows the (electron) density and temperature as a function of position and time, the synthesized GOES light-curve (overlaid on the observed one), and the calculated RHESSI source heights (overlaid on the observed values).  The GOES light-curve is in approximate agreement with the observed one, in terms of the maximum flux and cooling time, in both GOES passbands.  One feature worth noting is that the non-thermal emission comprises a non-negligible part of the total emission in both GOES channels.  The predicted RHESSI source heights show that the emission in the three wavebands of interest forms near the apex, due to thermal emission from pre-heating.  Since the loop is initially dense, the electrons lose a significant portion of their energy in the corona, driving a strong down-flow and thermal conduction front.  Thus, the thermal emission is seen to move downwards (particularly in the 3-6 keV channel).  The non-thermal emission dominates in the higher energy channels at early times, so that those sources are found at chromospheric depths.  As the chromosphere heats up, though, material begins to evaporate into the corona, where the increase in temperature up to 20\,MK along with the material flow causes an increase in thermal emission (compare the temperature and density plots).  In all three channels, therefore, the sources are observed to rise towards the apex of the loop.  Since the apex is the hottest part of the loop at late times and since the density is approximately constant across the corona, the emission is brightest at the apex (and roughly symmetric about it), so that the source heights remain there during the cooling phase.
\begin{figure*}[!t]
\centering
\begin{minipage}[b]{0.32\linewidth}
\centering
\includegraphics[width=\textwidth]{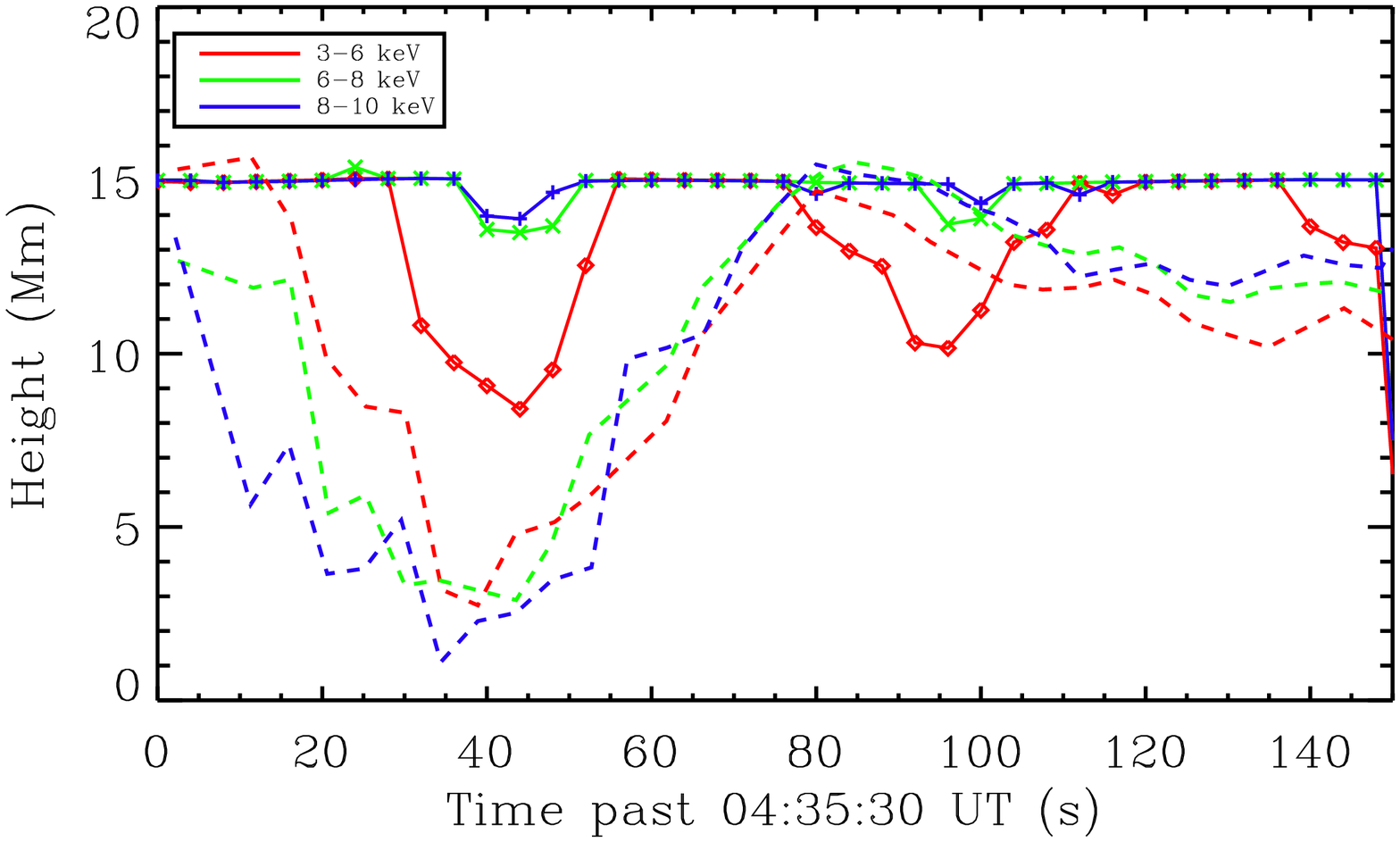}
\end{minipage}
\begin{minipage}[b]{0.32\linewidth}
\centering
\includegraphics[width=\textwidth]{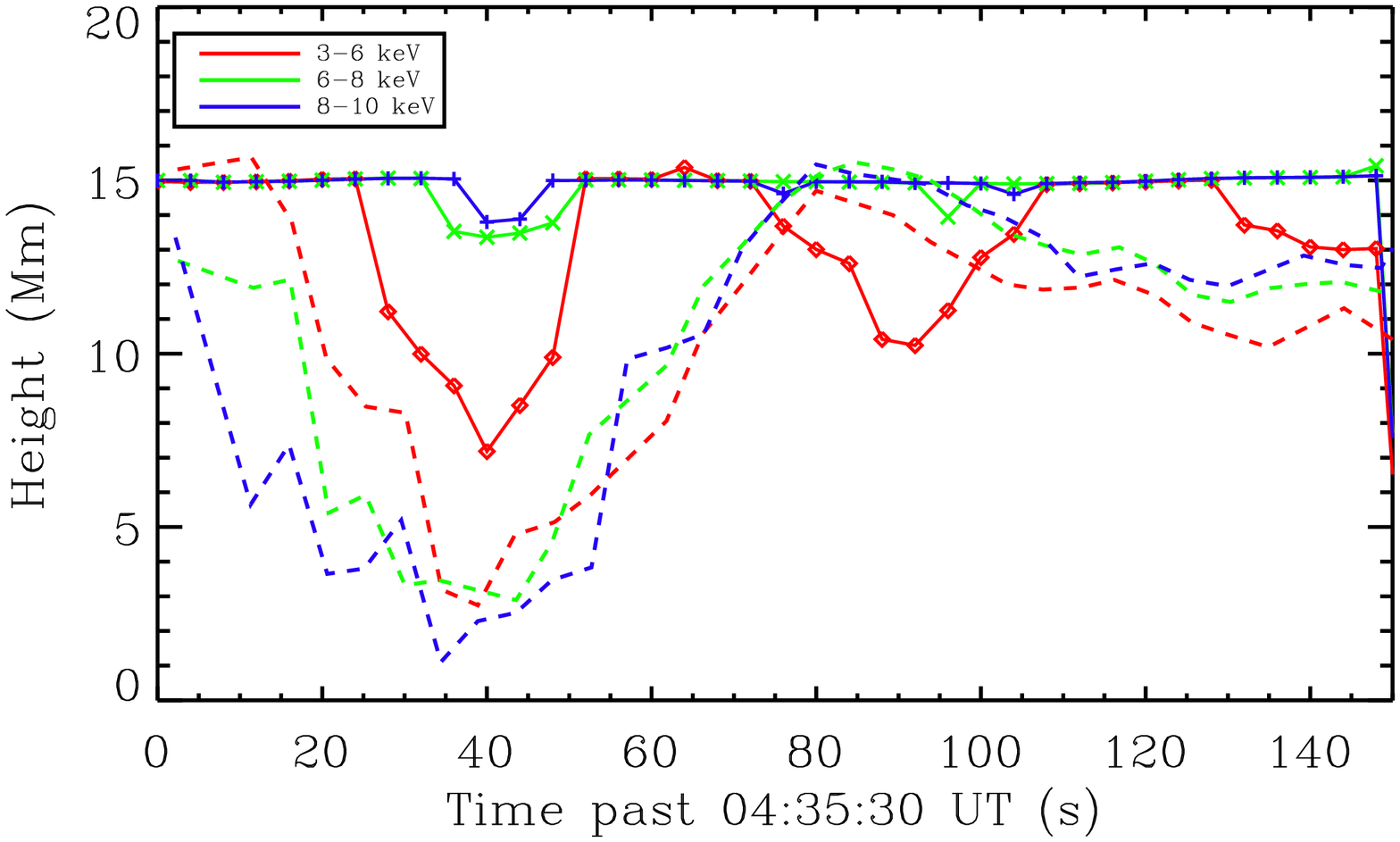}
\end{minipage}
\begin{minipage}[b]{0.32\linewidth}
\centering
\includegraphics[width=\textwidth]{run15_rhessi_heights.eps}
\end{minipage}
\begin{minipage}[b]{0.32\linewidth}
\centering
\includegraphics[width=\textwidth]{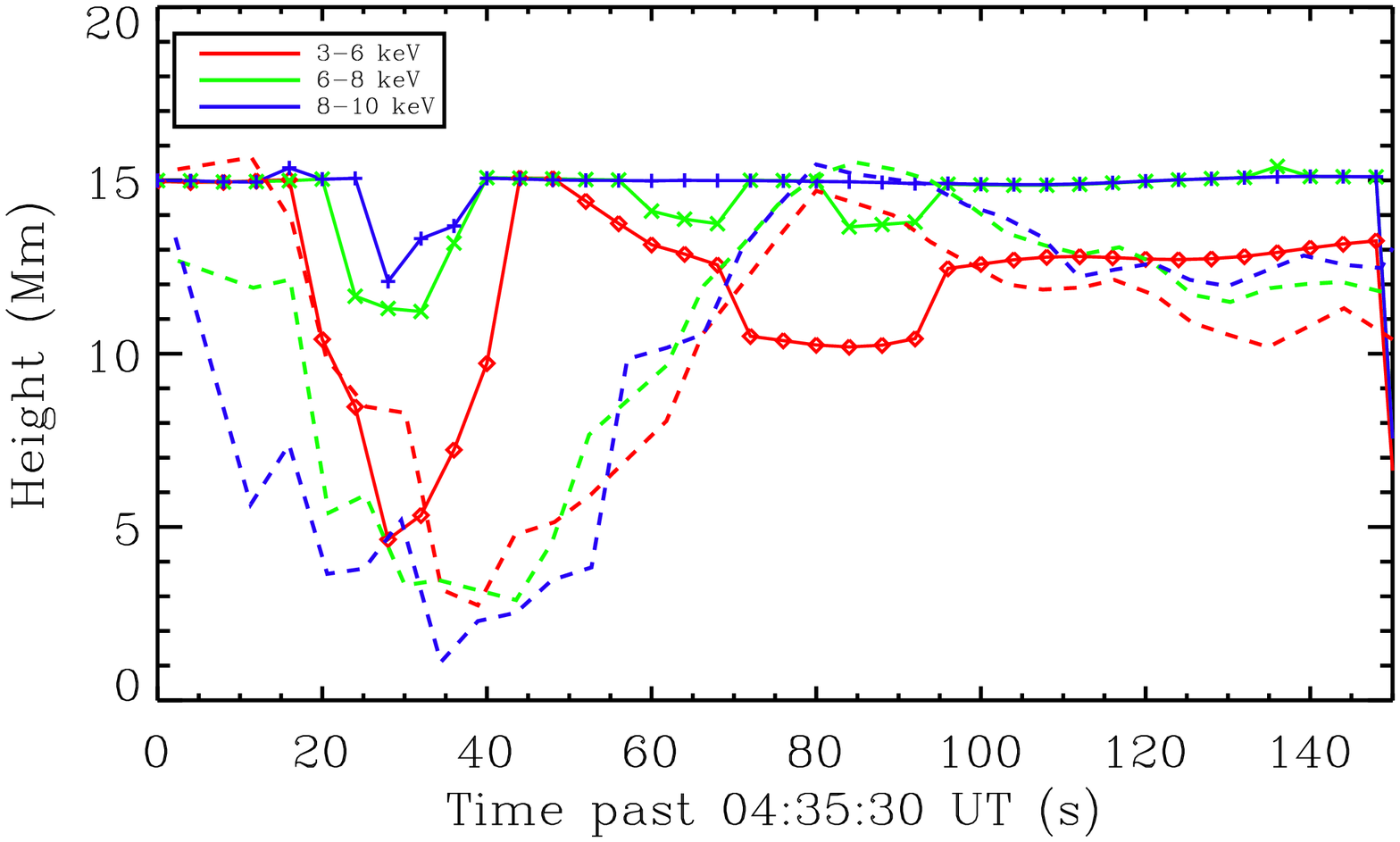}
\end{minipage}
\begin{minipage}[b]{0.32\linewidth}
\centering
\includegraphics[width=\textwidth]{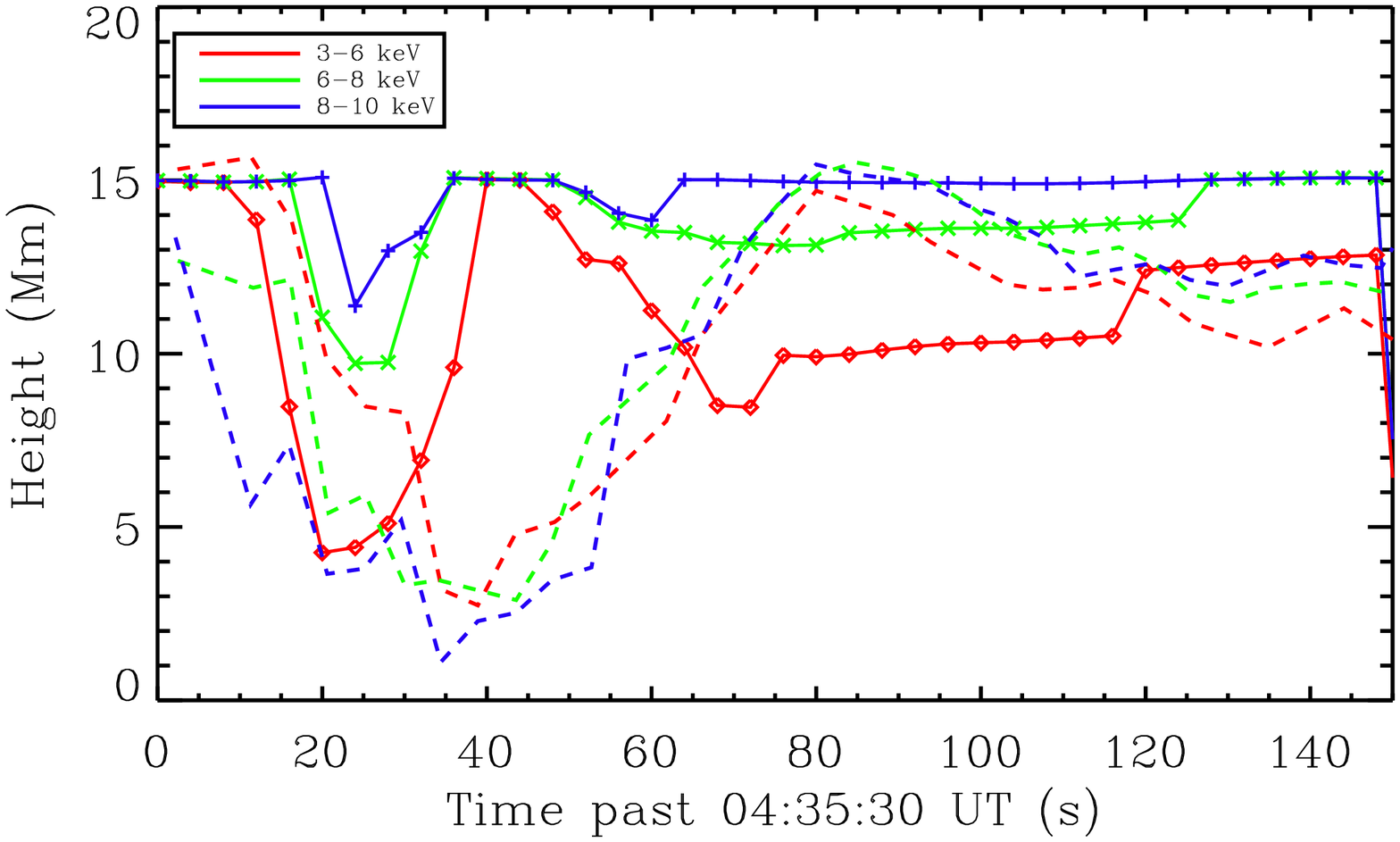}
\end{minipage}
\begin{minipage}[b]{0.32\linewidth}
\centering
\includegraphics[width=\textwidth]{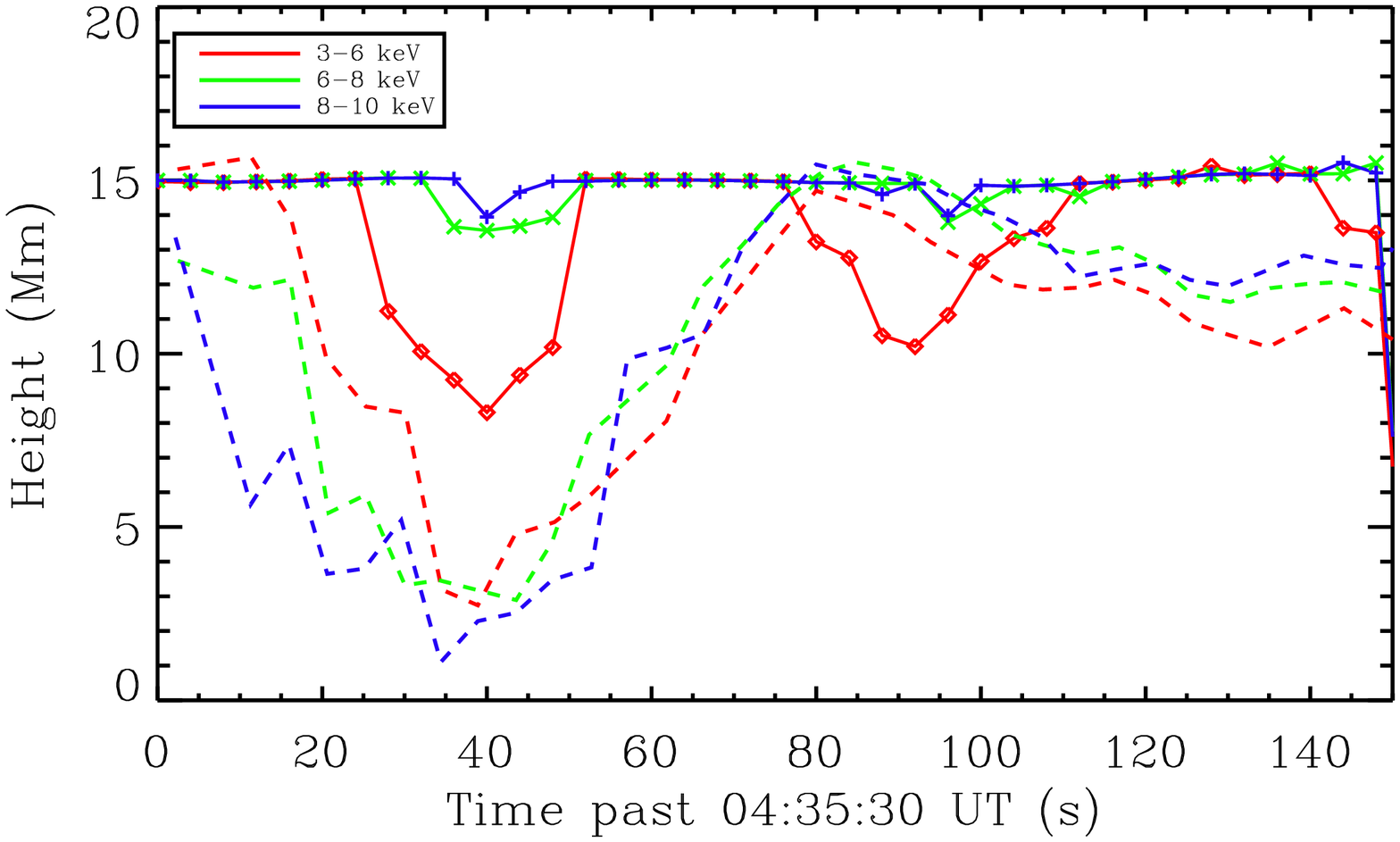}
\end{minipage}
\begin{minipage}[b]{0.32\linewidth}
\centering
\includegraphics[width=\textwidth]{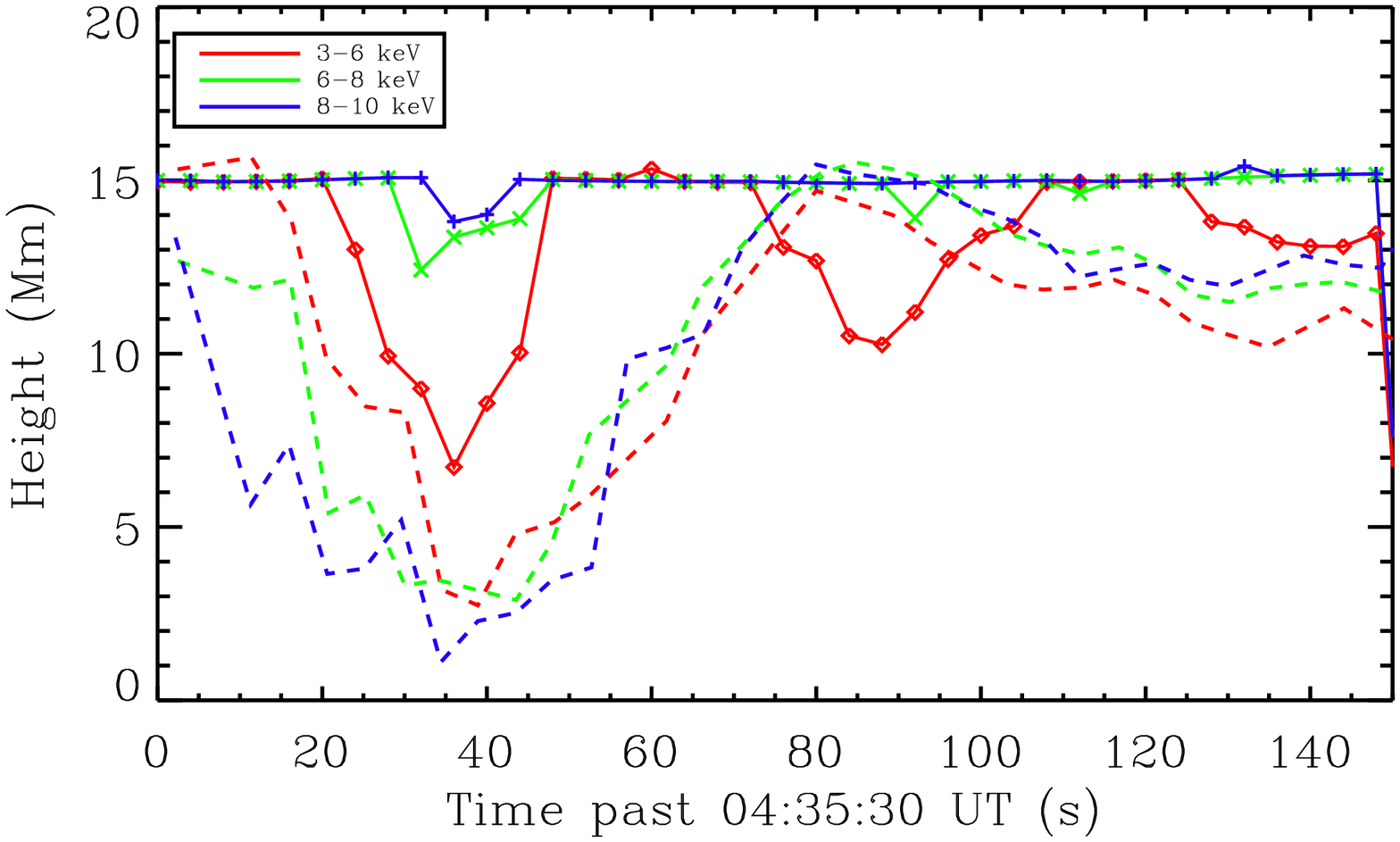}
\end{minipage}
\begin{minipage}[b]{0.32\linewidth}
\centering
\includegraphics[width=\textwidth]{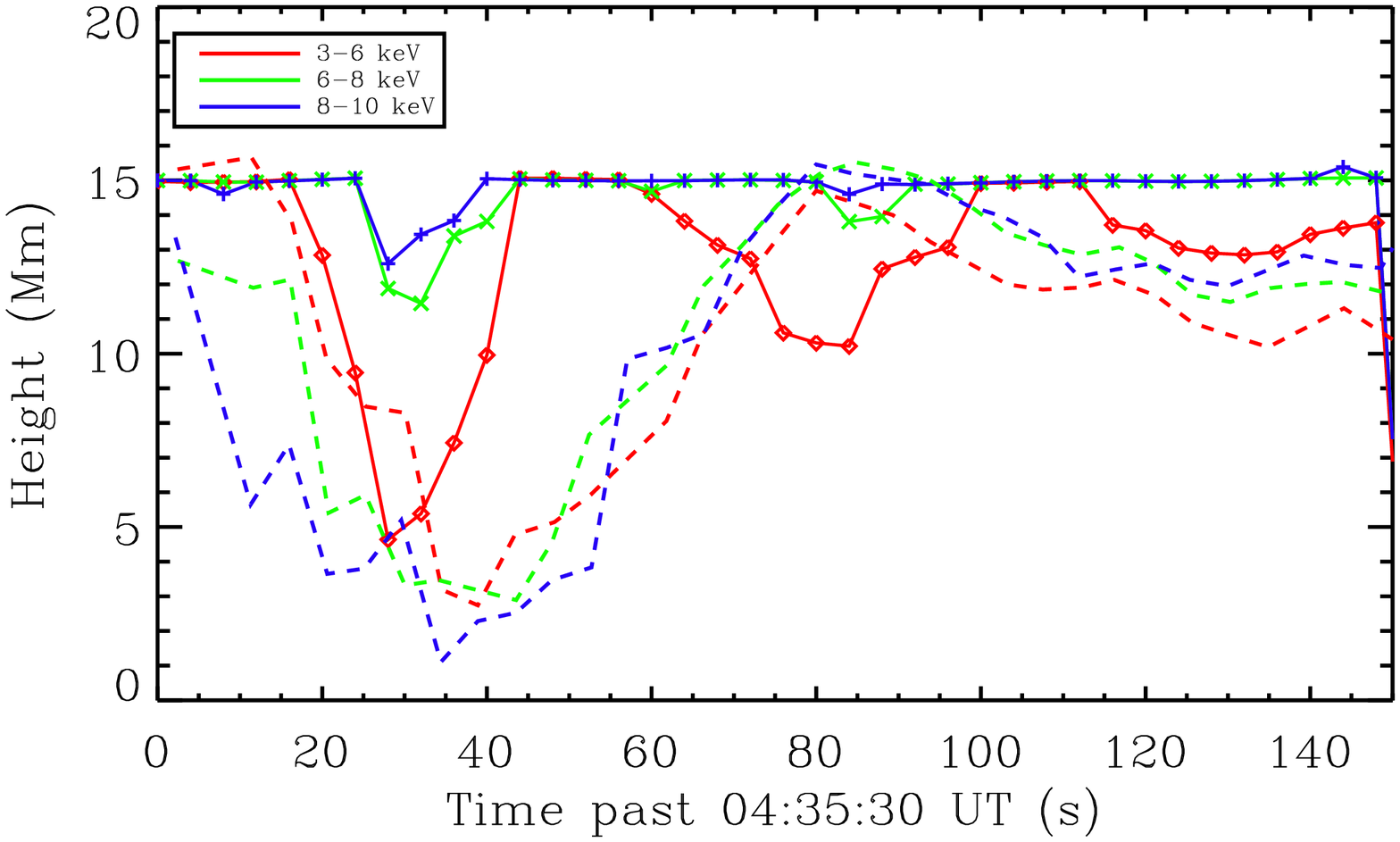}
\end{minipage}
\begin{minipage}[b]{0.32\linewidth}
\centering
\includegraphics[width=\textwidth]{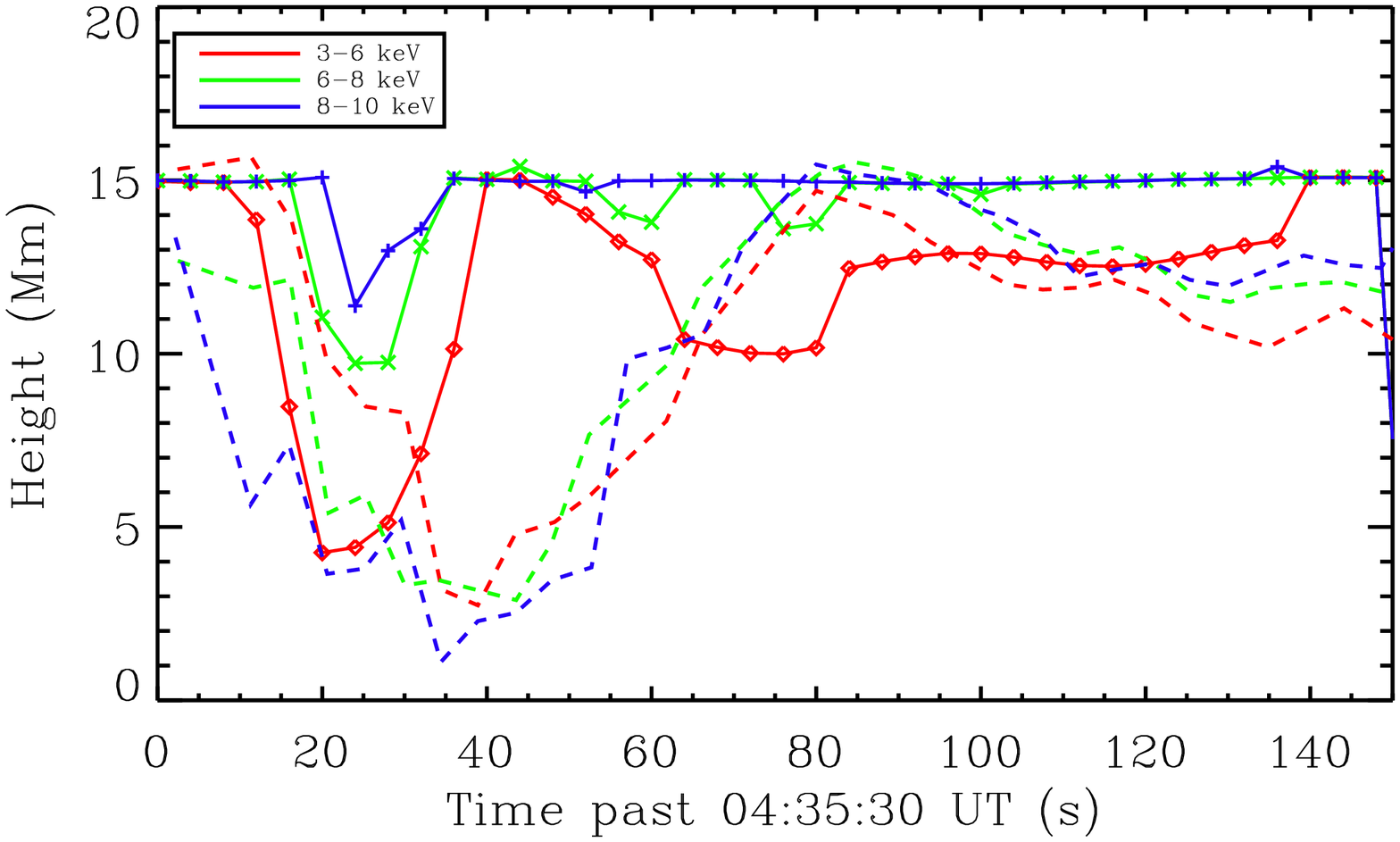}
\end{minipage}
\begin{minipage}[b]{0.32\linewidth}
\centering
\includegraphics[width=\textwidth]{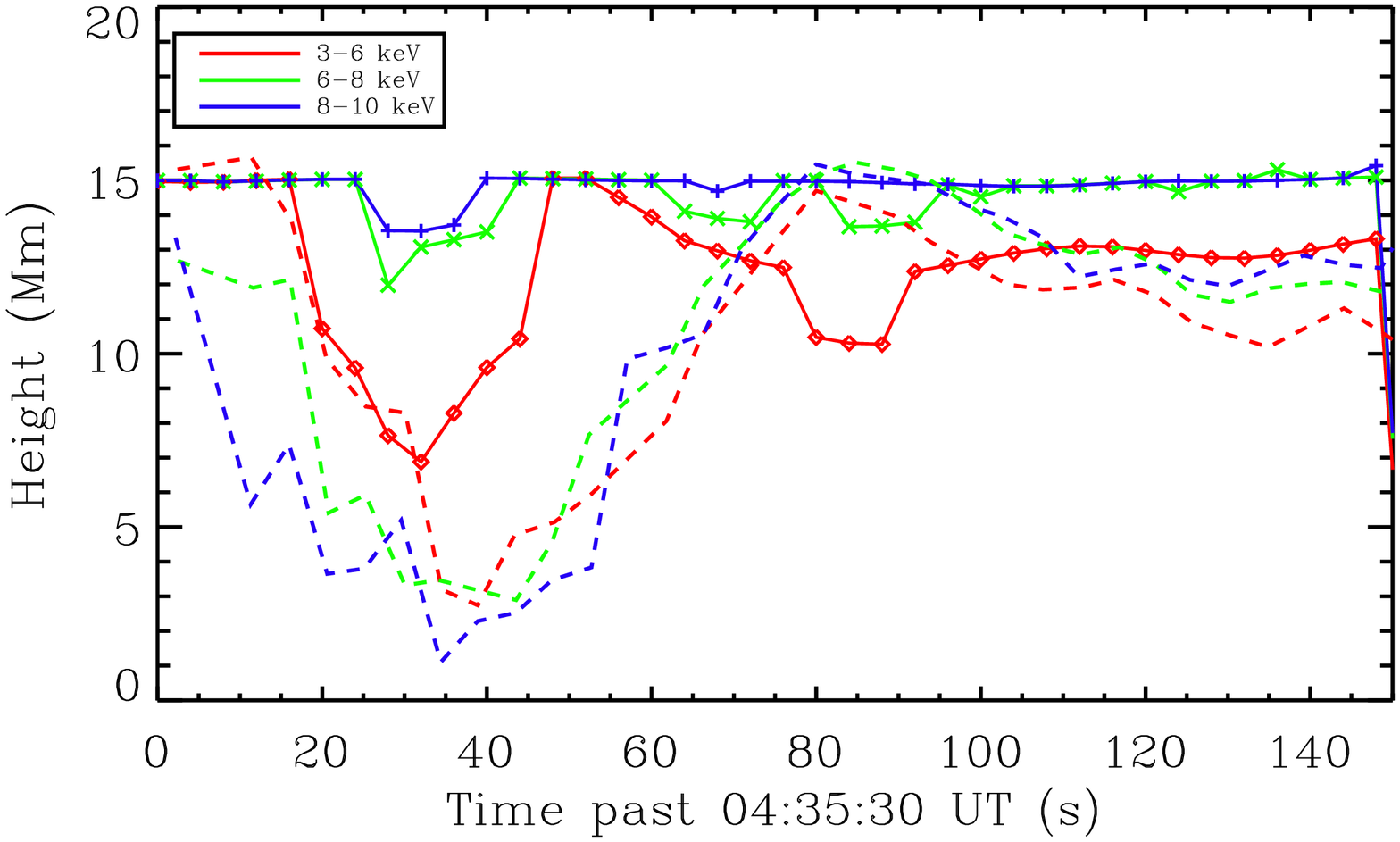}
\end{minipage}
\begin{minipage}[b]{0.32\linewidth}
\centering
\includegraphics[width=\textwidth]{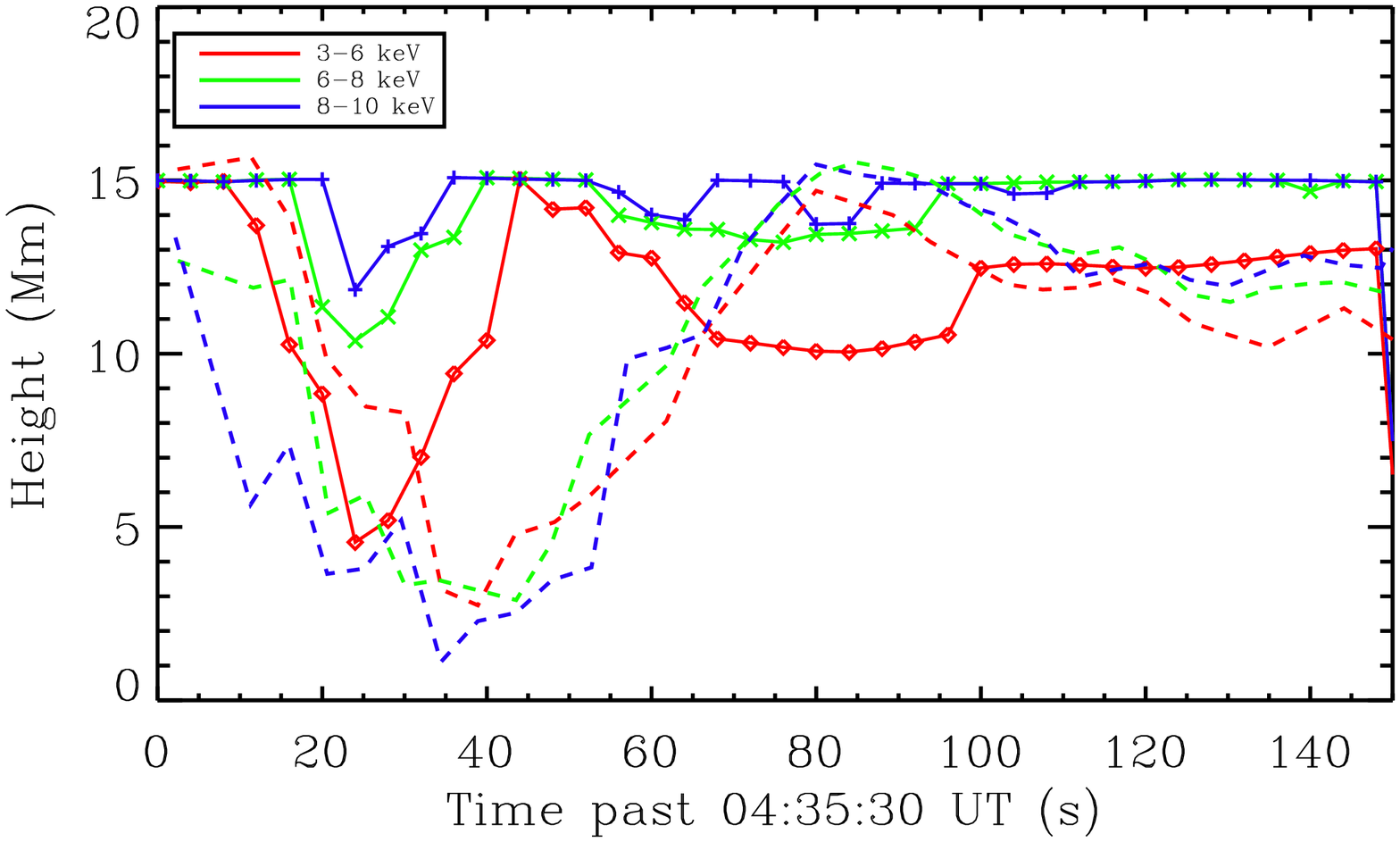}
\end{minipage}
\begin{minipage}[b]{0.32\linewidth}
\centering
\includegraphics[width=\textwidth]{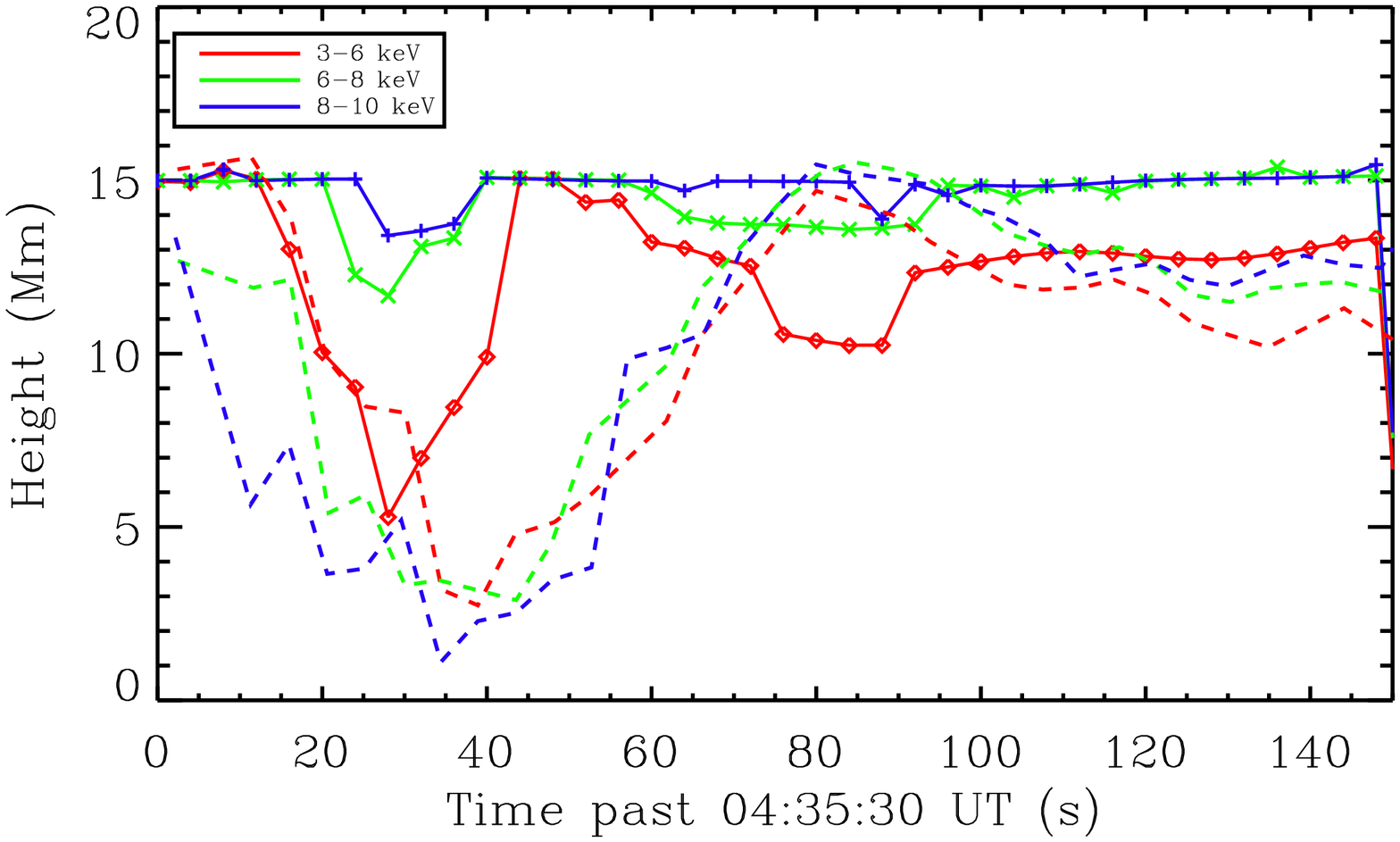}
\end{minipage}
\caption{The source heights calculated for the thermal models, Runs 13-24, from left to right and from top to bottom.  The timings, depths, and dispersion by energy are all inconsistent with the observed trends.}
\label{thermalheightshigh}
\end{figure*}

Next, consider a thermal model, Run 15, with {\it in situ} heating depositing a total of $5.0 \times 10^{27}$\,erg.  Figure \ref{run15} similarly shows the density, temperature, GOES flux, and calculated source heights as functions of time.  The heat is deposited over the top 1\,Mm of the loop, which quickly drives a thermal conduction front down the loop towards the lower atmosphere.  Around the transition region/chromosphere boundary, the energy dissipates, causing the top layers to boil off, which drives an up-flow of material.  The thermal emission becomes detectable almost immediately, centered around the apex of the loop.  As the thermal conduction front propagates downwards, the thermal emission in the lowest energy channel begins to fall.  However, the fall is significantly less drastic in the higher energy channels.  The emission in the 3-6 keV channel begins to rise as material from the chromosphere evaporates.  In this lower energy channel, the rise and fall times correspond very well with the observed sources.  The morphology of the higher energy sources, however, does not agree with the observations, as they do not depart significantly from the apex.  That the 3-6 keV source is lower than the others contradicts the observations, as well.
\begin{figure*}[!t]
\centering
\begin{minipage}[b]{0.32\linewidth}
\centering
\includegraphics[width=\textwidth]{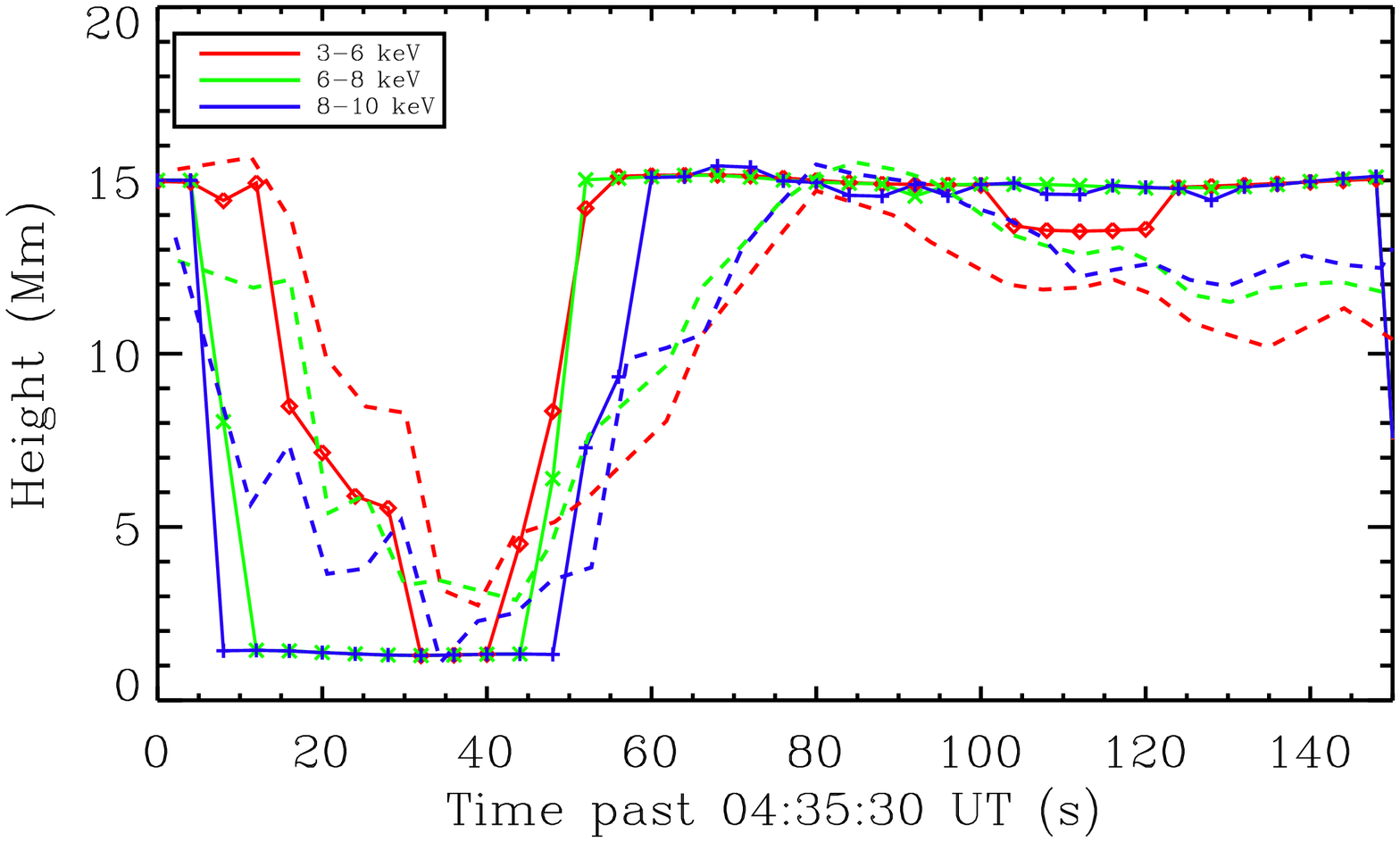}
\end{minipage}
\begin{minipage}[b]{0.32\linewidth}
\centering
\includegraphics[width=\textwidth]{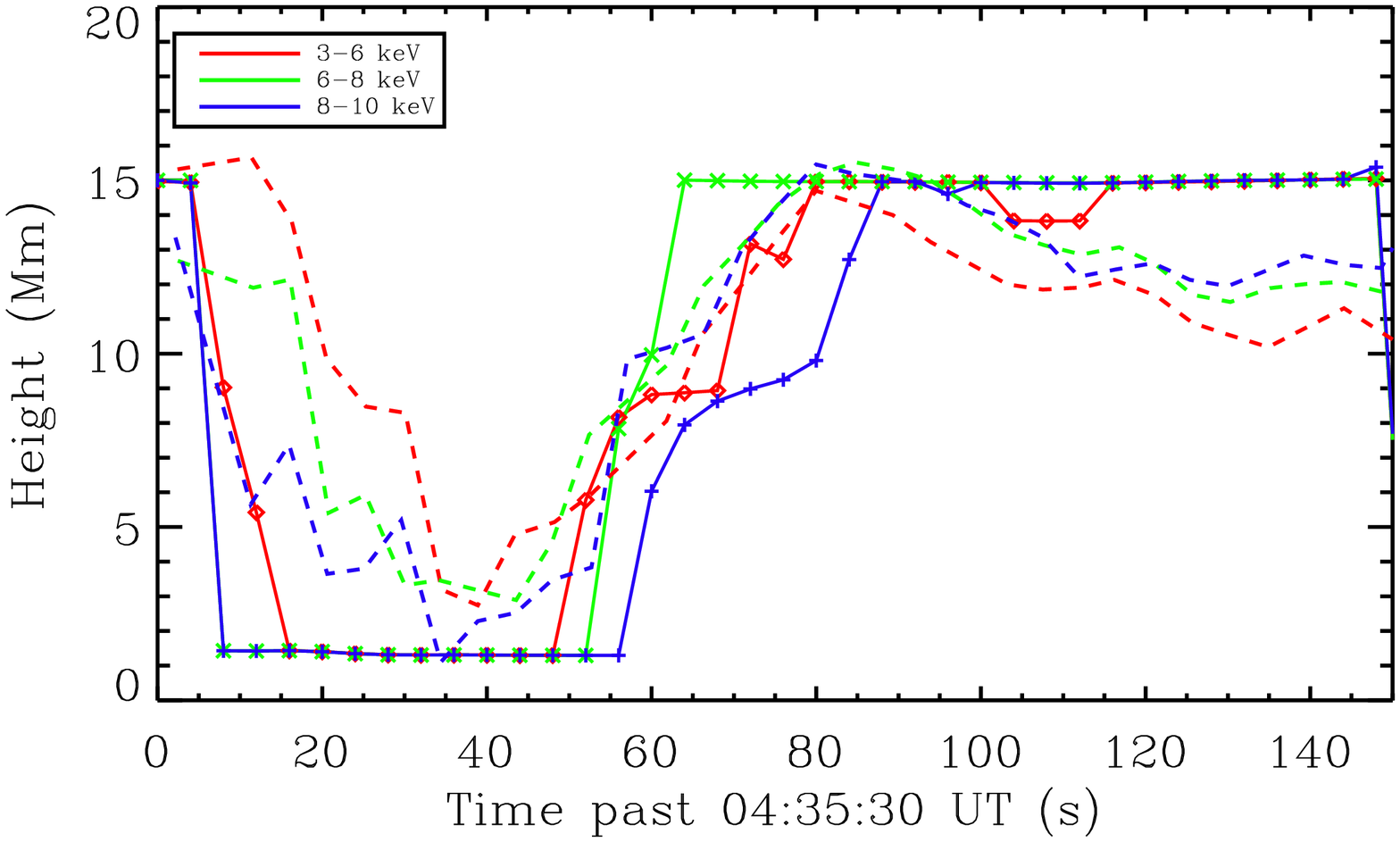}
\end{minipage}
\begin{minipage}[b]{0.32\linewidth}
\centering
\includegraphics[width=\textwidth]{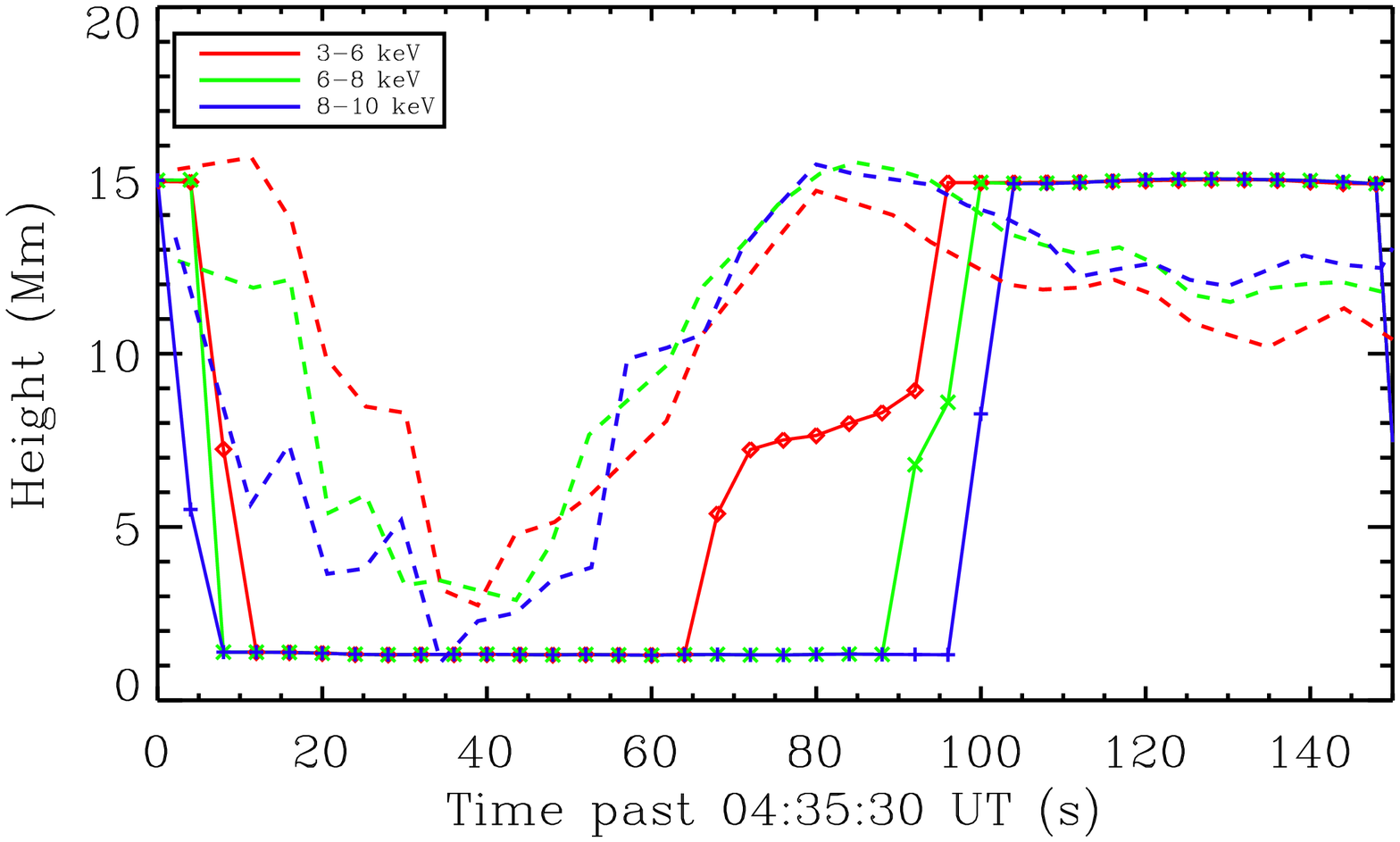}
\end{minipage}
\begin{minipage}[b]{0.32\linewidth}
\centering
\includegraphics[width=\textwidth]{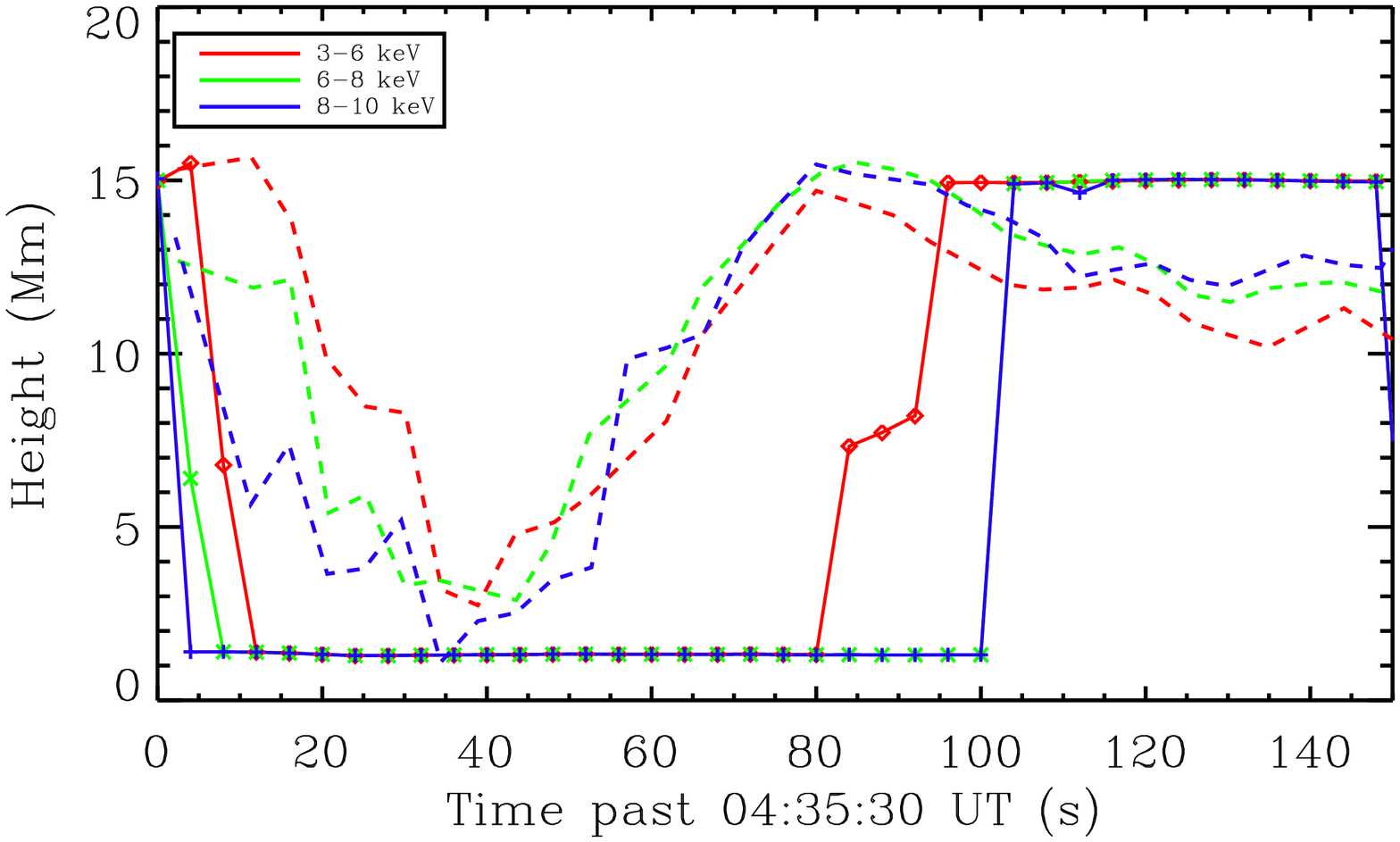}
\end{minipage}
\begin{minipage}[b]{0.32\linewidth}
\centering
\includegraphics[width=\textwidth]{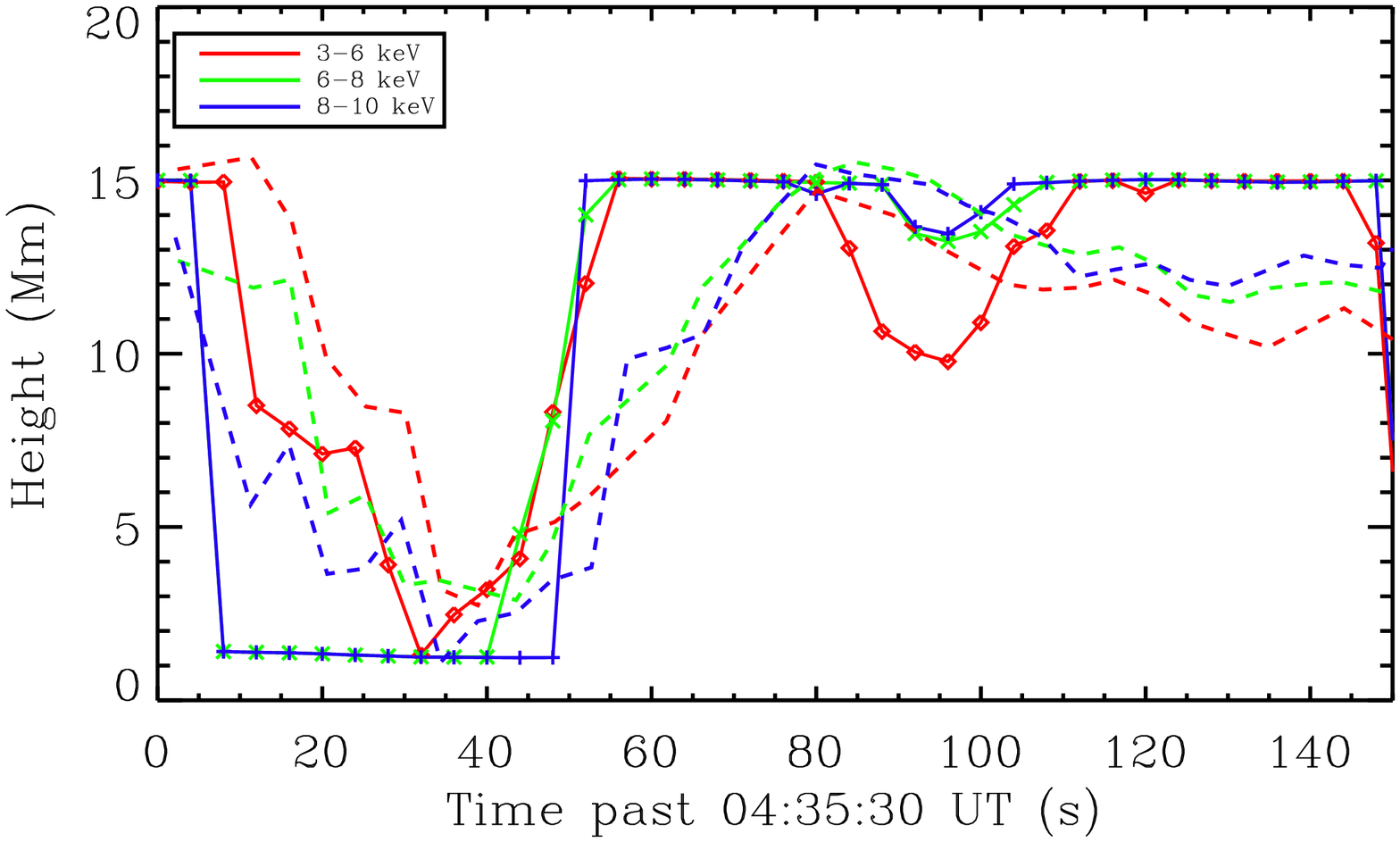}
\end{minipage}
\begin{minipage}[b]{0.32\linewidth}
\centering
\includegraphics[width=\textwidth]{run6_rhessi_heights.eps}
\end{minipage}
\begin{minipage}[b]{0.32\linewidth}
\centering
\includegraphics[width=\textwidth]{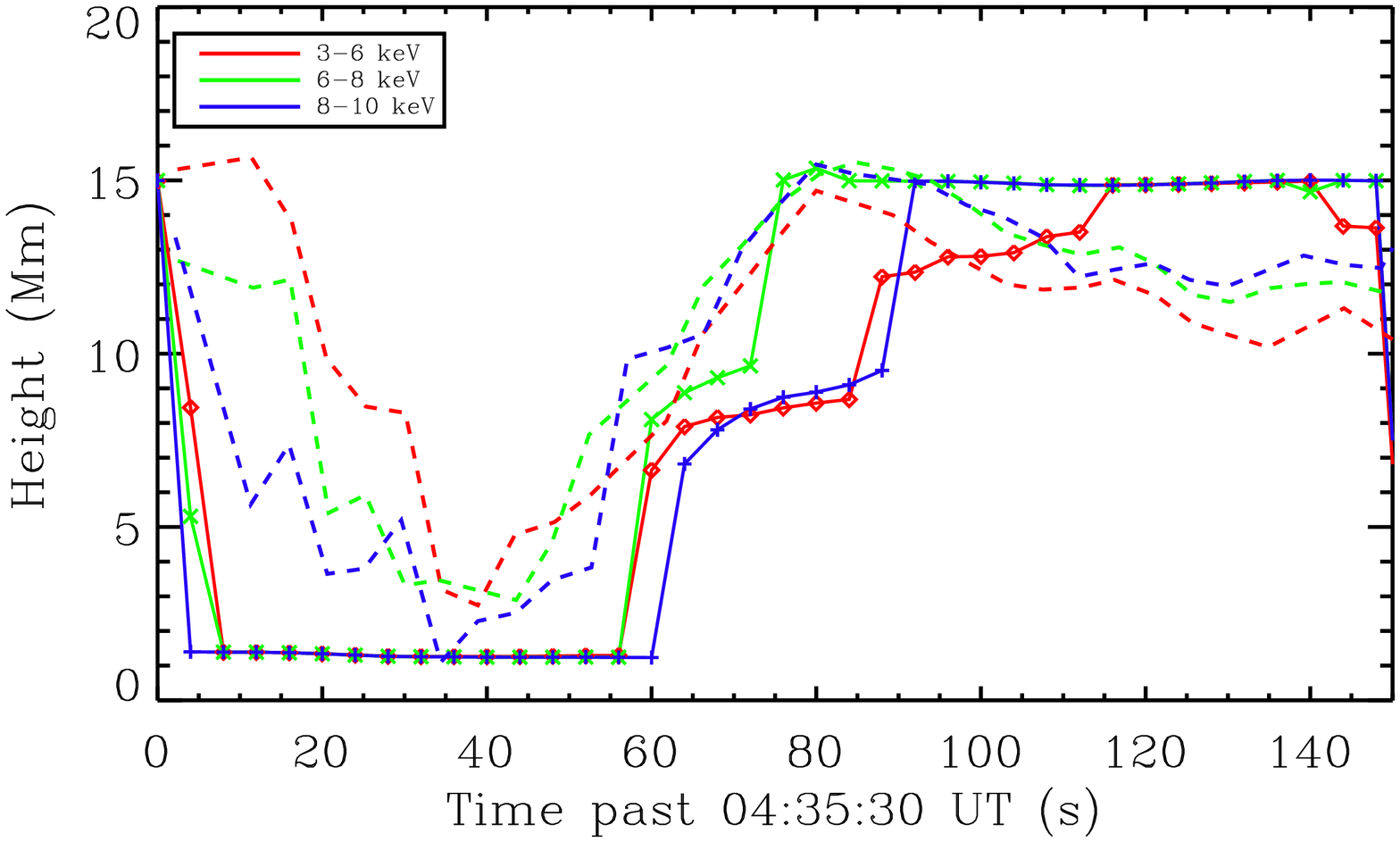}
\end{minipage}
\begin{minipage}[b]{0.32\linewidth}
\centering
\includegraphics[width=\textwidth]{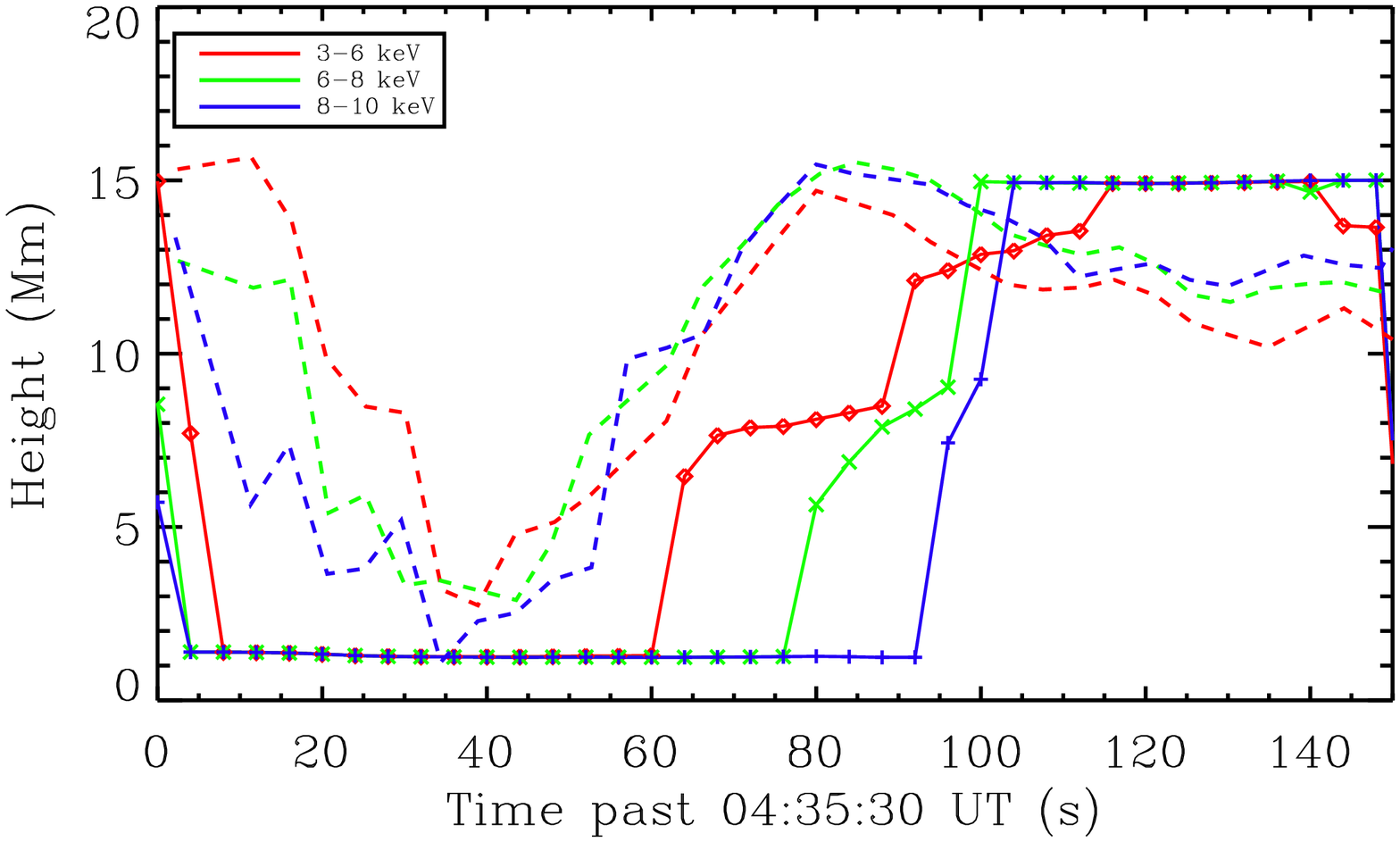}
\end{minipage}
\begin{minipage}[b]{0.32\linewidth}
\centering
\includegraphics[width=\textwidth]{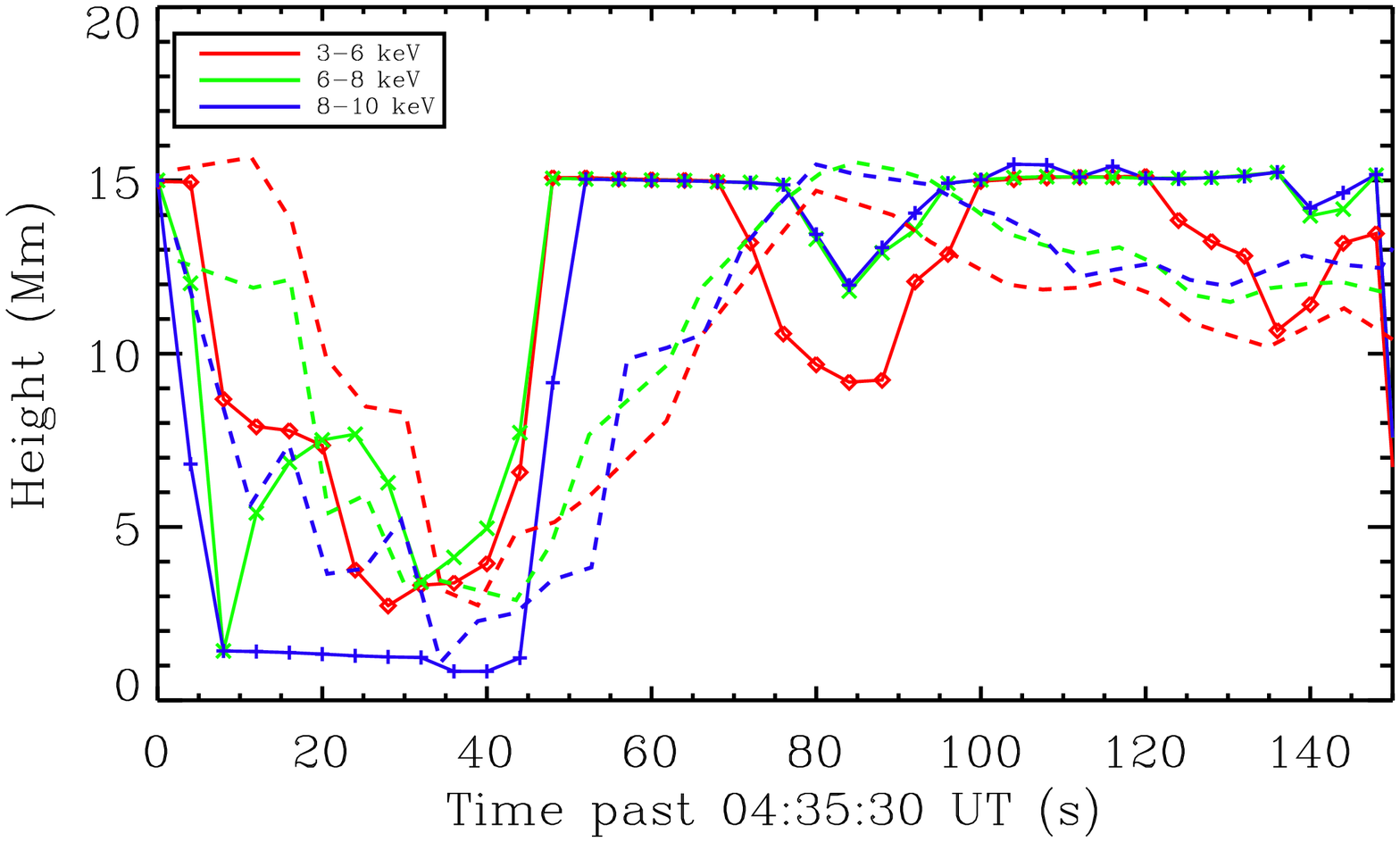}
\end{minipage}
\begin{minipage}[b]{0.32\linewidth}
\centering
\includegraphics[width=\textwidth]{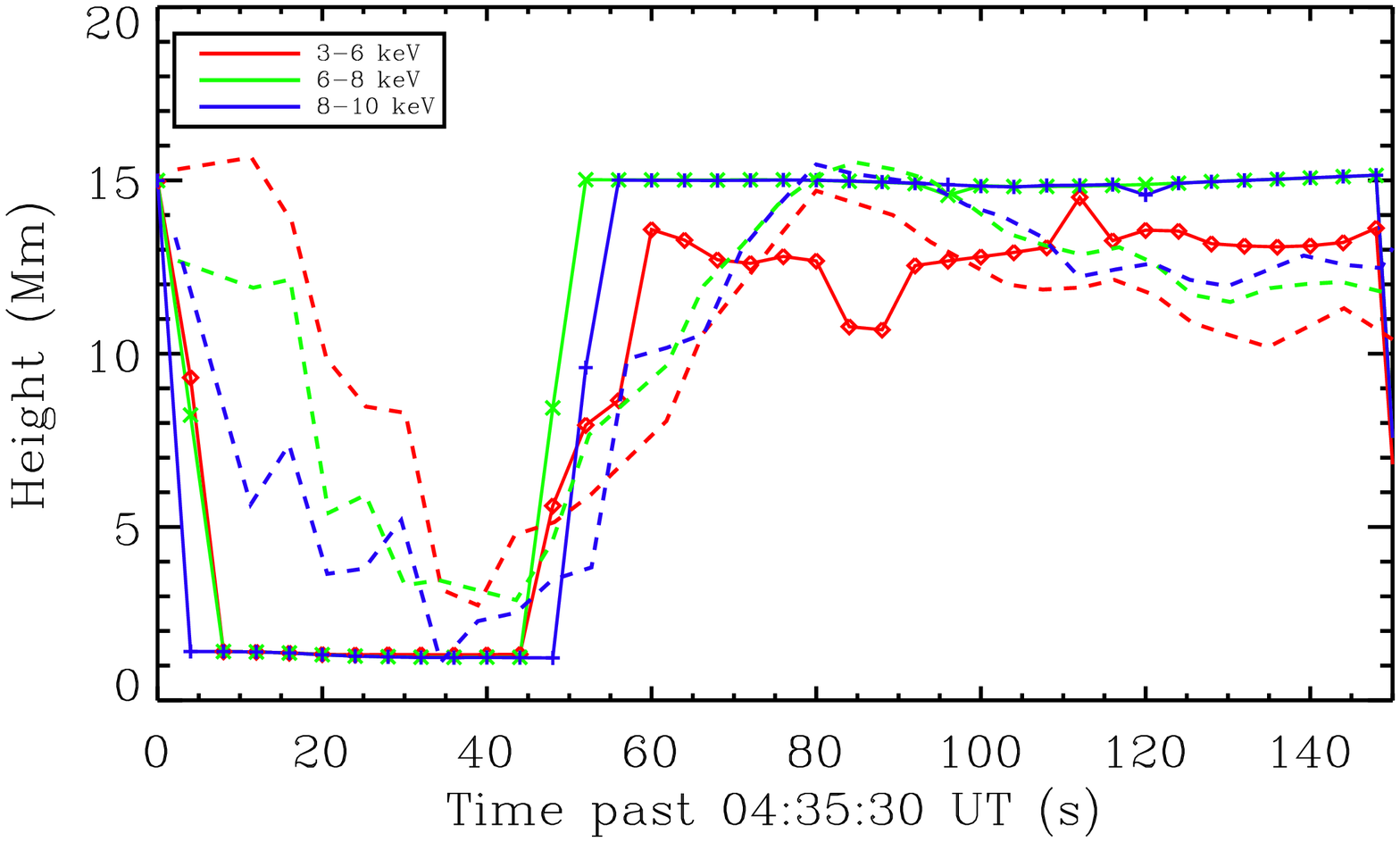}
\end{minipage}
\begin{minipage}[b]{0.32\linewidth}
\centering
\includegraphics[width=\textwidth]{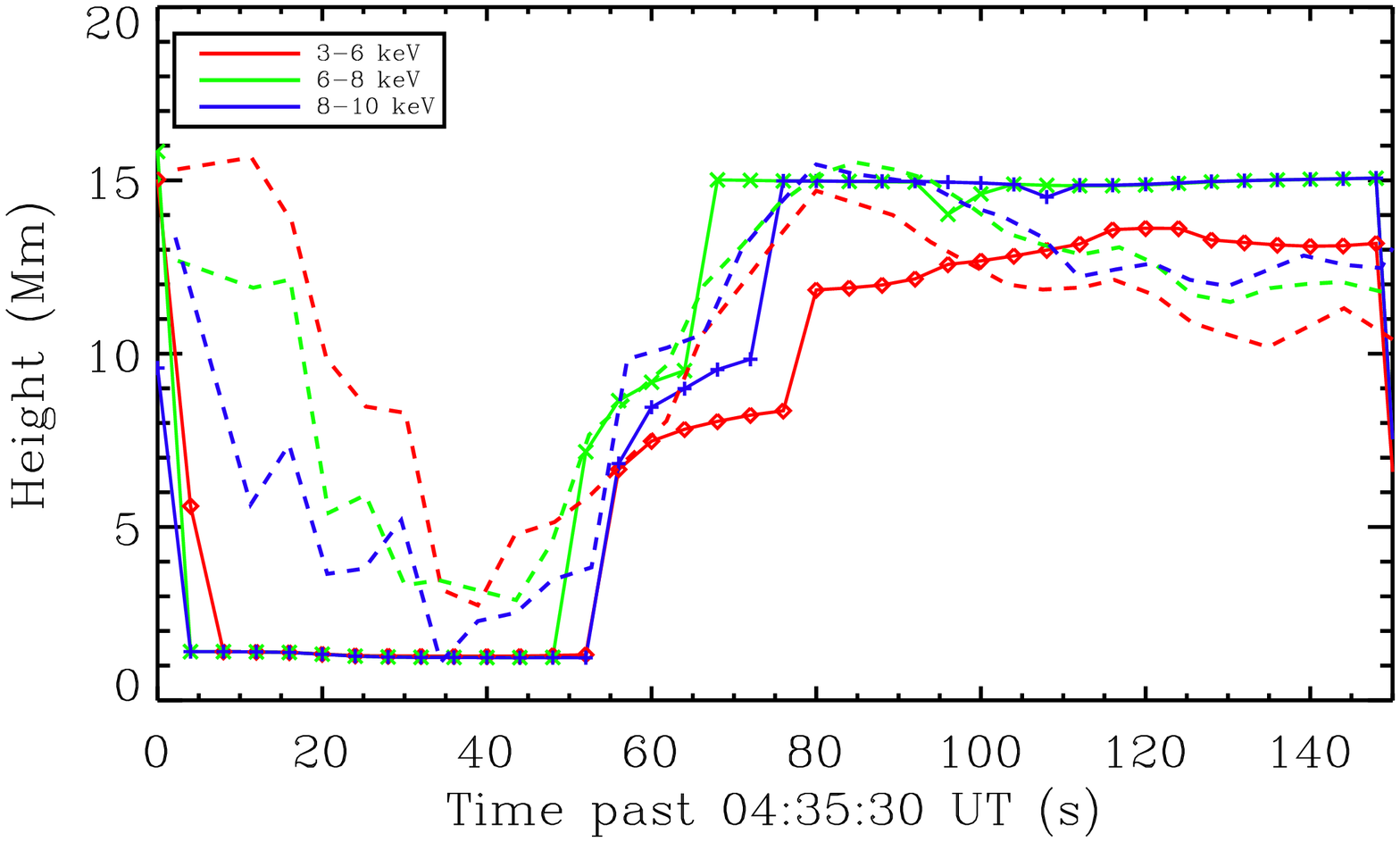}
\end{minipage}
\begin{minipage}[b]{0.32\linewidth}
\centering
\includegraphics[width=\textwidth]{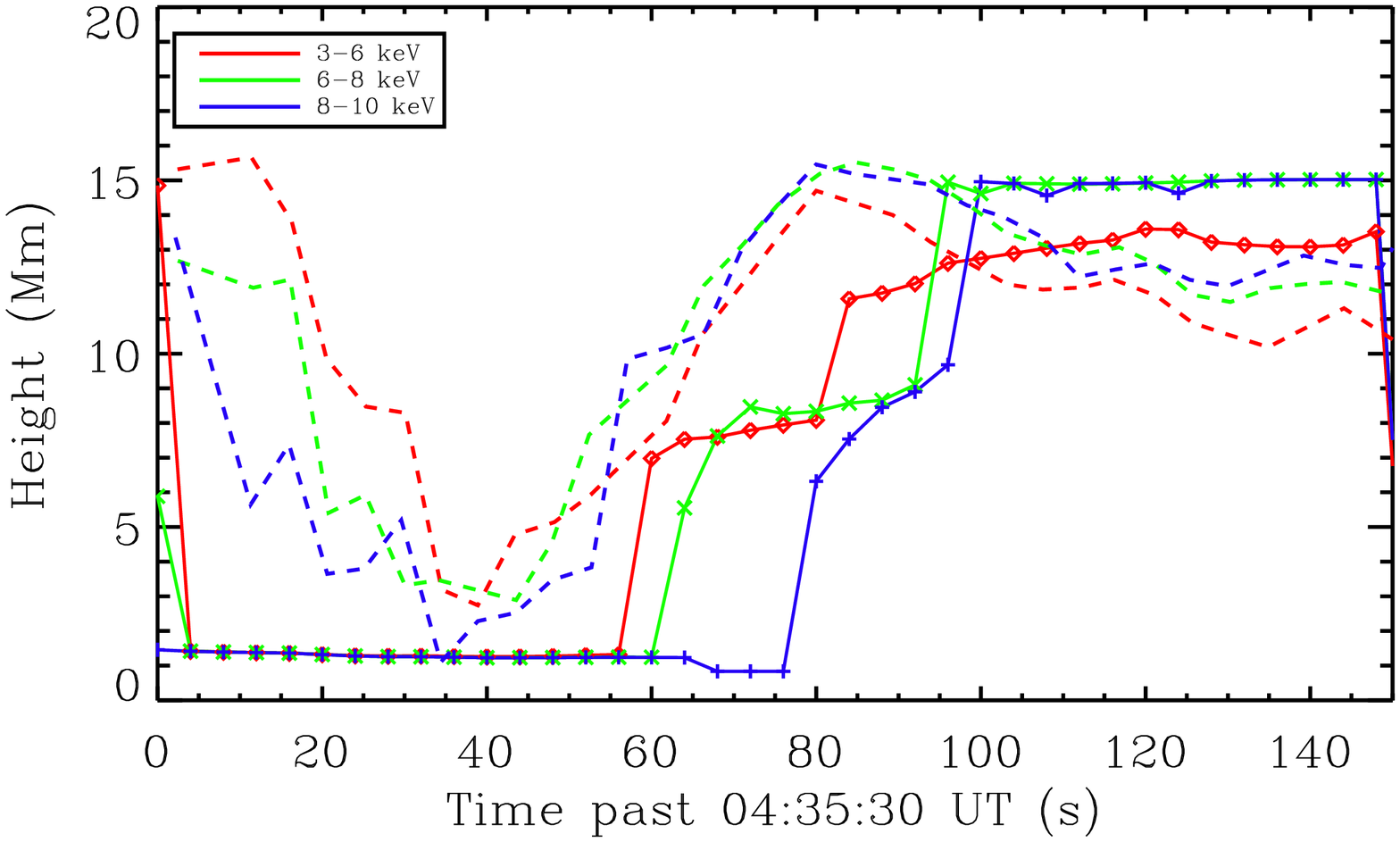}
\end{minipage}
\caption{The source heights calculated for Runs 1-12 (thick-target model), from left to right and top to bottom. }
\label{thickheightshigh}
\end{figure*}

All of the thermal simulations fail to reproduce the source height motion (Figure \ref{thermalheightshigh}).  Many of the simulations have reasonable matches in the low-energy 3-6 keV channel, in that they rise and fall at about the same time and to a similar depth as the observations.  The higher energy sources essentially remain at the apex throughout the simulations.  One other major problem with all of the thermal models is the height dispersion with energy of the sources.  The 3-6 keV source is always seen at lower heights than the higher energy ones, which is in contradiction with the heights during the impulsive phase of the flare.  The 6-8 keV source is generally lower than the 8-10 keV source during this time period, also in contradiction to the observed motions.  It is clear that a purely thermal heating model cannot reproduce the observed source motions.

Consider the beam heating models now (Figure \ref{thickheightshigh}).  The sources all begin at the loop apex, where there is a weak thermal source.  As the beam strengthens, non-thermal emission grows in the chromosphere, while thermal conduction fronts begin to form and move down the corona.  At later times, thermal emission dominates and the sources rise to the apex of the loop in all three channels.  In general, rising and falling motion is seen in all three energy bands.  The best match is perhaps Run 5, in terms of the timings and depths to which the sources fall.  Although none of the simulations are perfect matches, the reasonable agreement found suggests that the observations are consistent with a thick-target model.  Given the uncertainties involved, it would not be productive to tweak the parameter values to improve the match.  Finding reasonable agreement between the model predictions and the observed behavior with minimal intervention (particularly in terms of the overall timing and extent of the motion in separate energy channels) is sufficient.

\section{Summary \& Conclusions}
\label{conclusions}

24 numerical experiments of the C1.1 flare of 28 November, 2002 (04:35 UT) have been run.  Using initial temperatures and densities consistent with the derivations of \citet{oflannagain2013}, 12 simulations were performed using a thick-target model.  12 simulations were performed under the assumption of {\it in situ} coronal heating, distributed across length scales of [1, 10, 30] Mm.  All 24 simulations were found to be in approximate agreement with the observed GOES light-curves, although there are significant differences between the predicted motions of the RHESSI source heights and their observed morphology.  

In the thick-target models, due to the dense initial state of the corona, the electrons deposit a significant amount of their energy in the corona, although a large amount still penetrates to the chromosphere.  During the initial phase of heating, strong down-flows form, that, along with the thermal conduction front, carry material and energy from the corona to the chromosphere.  As the pressure builds in the chromosphere, evaporation brings material back into the corona.  The observed X-ray source motion can thus be described as the combination of a few processes.  At early times, the thermal source in the corona dominates, while the non-thermal emission in the chromosphere grows.  As the spectrum hardens so that there are more electrons at higher energies, and as material flows downwards, the sources fall.  At later times, as material evaporates, the emission thermalizes, and the sources rise to the apex, where they remain throughout the cooling phase.  The observations are consistent with the collisional thick-target model.

In the purely thermal models, however, the source motion cannot be reproduced.  Strong heating, when confined near the apex of the loop, produces a sharp rise in the coronal temperature to nearly 30\,MK, which similarly drives a strong conduction front.  Because thermal emission quickly drops off at higher photon energies, the higher energy bands are more strongly impacted by the absence of non-thermal bremsstrahlung.  These sources are most intense at the apex, where the temperature is the highest.  Away from the apex, due to the large gradient in the temperature, there is little emission at these energies.  Therefore, for a range of heating parameters, the 6-8 and 8-10 keV channels were found to not depart significantly from the apex of the loop.  Further, the synthesized 3-6 keV source was found at lower heights than the other two energy bands in all of the simulations, and the 6-8 keV source was lower than the 8-10 keV source, which contradicts the observations.  Although a purely thermal model can reproduce the GOES light-curves, it cannot simultaneously reproduce the observed source motions, and we must conclude that the observations are inconsistent with it.

There are a few caveats.  First, in these simulations the cut-off energy is extremely low ($< 7$\,keV).  Both \citet{sui2006} and \citet{oflannagain2013} note that the non-thermal emission in this flare extends down to the threshold of RHESSI's detection (about 3 keV), and presumably beneath it.  While this cut-off energy is low, it is not implausible.  Second, the loop must have been preheated for consistency with the observations.  The initial coronal density used in the simulations here, however, is consistent with the findings of O'Flannagain et al. (see their Figure 5) and with the general properties of flaring active regions.  Third, we do not know the orientation of the flaring loop with respect to vertical, or whether this flare occurred on multiple strands, which could impact the observed properties.

As noted by O'Flannagain et al., pitch angle diffusion of the electron beam could also be an important factor in the observed heights.  The electron distribution could have a high pitch angle ({\it e.g.} more perpendicular to the field line), which gradually becomes more isotropized via collisions, allowing a greater proportion of accelerated electrons to propagate further and further down the loop in time.  \citet{winter2011} provide evidence that pitch angle diffusion could affect the observed source heights, which needs to be examined in more depth.

By combining the observational results with detailed hydrodynamic simulations, the method by which energy is released in a small flare has been examined in detail.  The flare was exceptional in that it occurred at the limb and was clearly observed with RHESSI, which was then at its peak sensitivity.  It was found that a beam heating scenario can reproduce observed X-ray source motions, although a purely thermal model cannot within the bounds of the numerical experiments performed here.  In order for a thermal model to be plausible, there needs to be a mechanism to emit high energy photons from lower heights than less energetic ones.  If similar observations are performed for other flares, whether with RHESSI or a new instrument such as FOXSI \citep{krucker2014}, this test can be repeated to further constrain the energy release mechanism and the numerical model.

\vspace{5mm}

This research was performed while JWR held an NRC Research Associateship award at the US Naval Research Laboratory with support from NASA, and previously by NASA Headquarters under the NASA Earth and Space Science Fellowship Program (Grant NNX11AQ54H).   GDH was supported by a NASA Living with a Star TR\&T Grant and the RHESSI Project.  The authors thank Giulio Del Zanna for his suggestions about evaluating X-ray emission with CHIANTI in our model.  CHIANTI is a collaborative project involving George Mason University, the University of Michigan (USA) and the University of Cambridge (UK).

\end{document}